\documentclass[sigconf,authorversion]{acmart}
\AtBeginDocument{%
  }

\copyrightyear{2026}
\acmYear{2026}
\setcopyright{cc}
\setcctype{by}
\acmConference[NordiCHI '26]{Proceedings of the 14th Nordic Conference on Human-Computer Interaction}{October 03--07, 2026}{Vaasa, Finland}
\acmBooktitle{Proceedings of the 14th Nordic Conference on Human-Computer Interaction (NordiCHI '26), October 03--07, 2026, Vaasa, Finland}
\acmDOI{10.1145/3829807.3829878}
\acmISBN{979-8-4007-2373-5/2026/10}

\usepackage{makecell}
\usepackage{subcaption}

\begin{document}

\title{Artly: Exploring Digital Artists' Perceptions of AI-Generated Feedback}

\author{Ulvi Rajabli}
\orcid{0009-0009-5896-1422}
\affiliation{%
  \institution{LMU Munich }
  \city{Munich}
  \country{Germany}
}\email{ulvi.rajabli@campus.lmu.de}

\author{Alexander Wiethoff}
\orcid{0000-0003-2938-7272}
\affiliation{%
  \institution{LMU Munich}
  \city{Munich}
  \country{Germany}}
  \email{alexander.wiethoff@ifi.lmu.de}

\author{Zelun Tony Zhang}
\orcid{0000-0002-4544-7389}
\affiliation{%
  \institution{TU Munich}
  \city{Munich}
  \country{Germany}}
\affiliation{
\institution{Munich Center for Machine Learning (MCML)}
  \city{Munich}
  \country{Germany}
}
\email{zelun.tony.zhang@tum.de}

\begin{abstract}
Recent developments in generative AI have lowered barriers to image generation, but existing tools mostly optimize for efficiency, producing generic results and offering little support for artistic growth. We present \textit{Artly}, an AI system that combines personalizable AI feedback with human-authored learning resources. In a between-subjects study with artists, we compared a mode without image generation features against one that allowed to generate variations of users' illustrations. \textit{Artly} was perceived as helpful for learning and self-improvement, with the exception of the most proficient participants. Participants who used the image generation feature interacted slightly less with the AI feedback. They reported feeling more creative after using \textit{Artly} than participants using the restricted mode, while reporting slightly lower scores on new ideas for their work. Overall, our findings underline the potential of our feedback approach for supporting artistic growth in a manner that is well received by artists.
\end{abstract}


\begin{teaserfigure}
  \includegraphics[width=\textwidth]{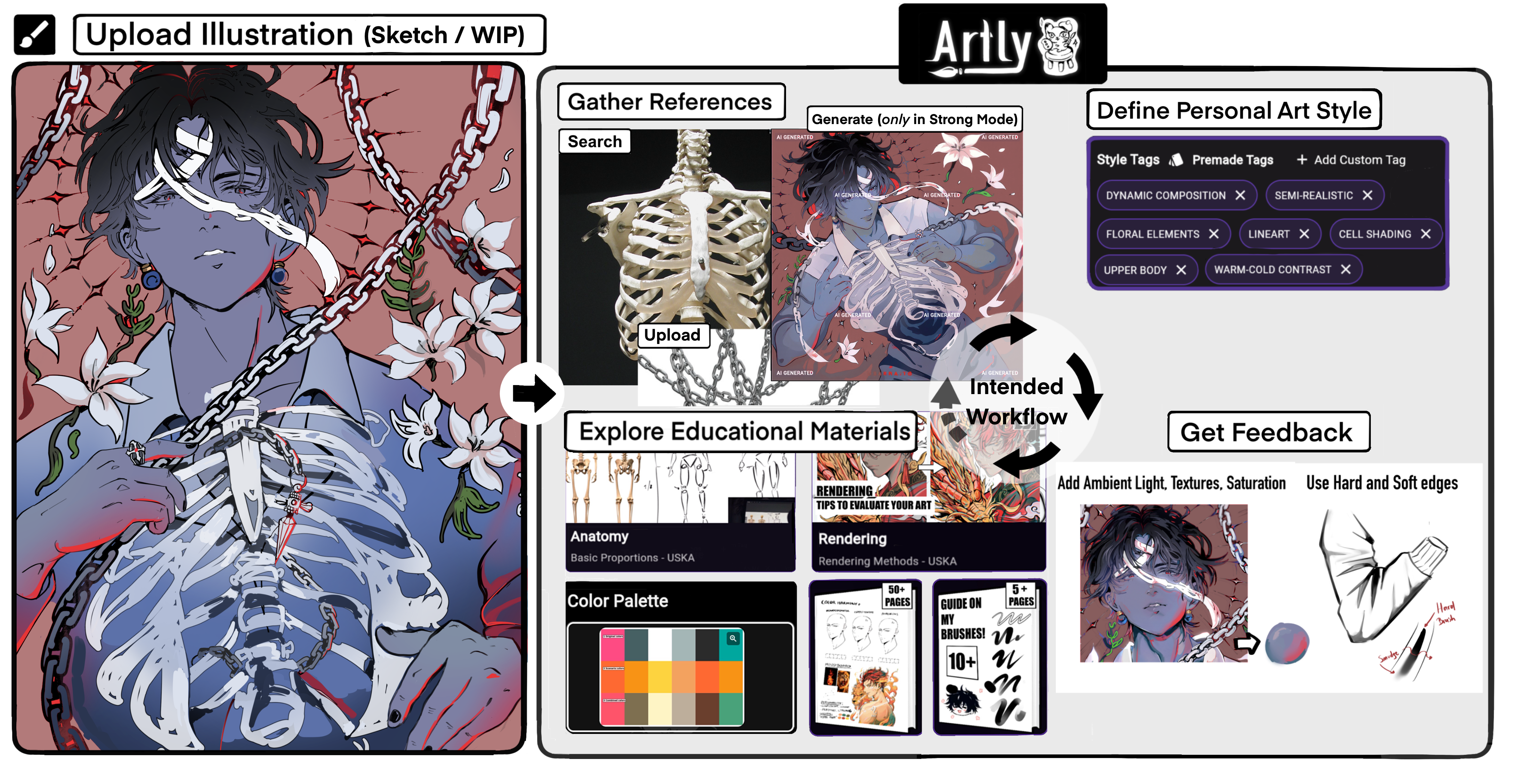}
  \caption{Overview of \textit{Artly}'s user workflow. Users start by uploading their illustration and gathering references. References can be uploaded, searched, or---as the defining feature of the Strong Mode variant---generated. 
  \textit{Artly} suggests style tags extracted from the references. Users can modify the suggested tags or add their own custom tags. These style tags define users' personal style and steer the AI-generated feedback. In addition to the AI feedback, users can explore additional resources including educational tutorials, brushes, and color palettes.}
  \Description{A diagram illustrating the "Artly" workflow system. On the far left, an initial panel titled "Upload Illustration (Sketch / WIP)" shows a digital illustration of a character with a ribcage motif and chains. An arrow points from this input into the main "Artly" container. The central "Artly" workspace consists of four panels arranged in a quadrant, connected by a central circular arrow labeled "Intended Workflow." The panels are: (1) "Gather References" (top-left), showing a 3D ribcage and chain references; (2) "Define Personal Art Style" (top-right), featuring style tags like "DYNAMIC COMPOSITION" and "CELL SHADING"; (3) "Get Feedback" (bottom-right), showing the original illustration with specific artistic critiques and technique notes; and (4) "Explore Educational Materials" (bottom-left), with resources for anatomy, color palettes, and rendering. The circular workflow indicates an iterative process between these four modules.}
  \label{fig:teaser}
\end{teaserfigure}

\maketitle

\section{Introduction}
The rapid development of generative AI technology has brought both new possibilities as well as significant challenges for artists and creatives. With generative AI, workflows tend to shift from detailed manual creation to orchestration and curation of AI outputs~\cite{palani2024evolving,tsaoPerceptionsIntegrationGenerative2025}. Creatives have to weigh benefits such as automation of repetitive tasks or fast generation of alternatives against challenges like the difficulty of articulating goals and lack of control over AI outputs~\cite{palani2024evolving}. Commercially oriented fields have been eager to adopt generative AI for its promised efficiency gains~\cite{tsaoPerceptionsIntegrationGenerative2025}. Especially novice designers tend to embrace it more enthusiastically, while senior designers tend to be more cautious and more concerned about erosion of traditional creative skills and expertise due to overreliance on AI~\cite{10.1145/3706598.3713233,tsaoPerceptionsIntegrationGenerative2025}.

Much of these concerns and challenges is driven by the circumstances that many widely used generative AI creativity tools, such as Midjourney, Adobe Firefly, or Leonardo.Ai, center on highly automated, prompt-based workflows. These tools turn text or image prompts into complete, high-fidelity candidate outputs with relatively little support for incremental control\footnote{While these tools increasingly offer options for control, they are mostly controls for editing or iterating on generated outputs. The primary workflow is still one-shot generation.} or skill development. Recent research on creativity support tools addresses this limitation by proposing various interaction designs where generative AI is used for more incremental support and where users remain more firmly in the loop. These tools are often either co-creation or ideation tools~\cite{Ko_2023_IUI_artists}: Co-creation tools increase the amount of human input while generating assets with AI. Ideation tools refrain from the production stage and instead help users to generate ideas for what to produce. 

However, tools proposed in both of these categories are mostly productivity-focused and, while under users' control, still take over a significant portion of the creative execution. This may not fit artistic contexts where the primary purpose is self-expression rather than productivity, and where artists may want to retain the low-level creation work. At the same time, artists have particularly deeply rooted aversions against generative AI in their profession due to concerns over job displacement, copyright infringements, lack of originality in AI outputs, and fear of being accused of using AI~\cite{kawakamiImpactGenerativeAI2024}.

We aimed to explore how generative AI can be used to support art-making for creative self-expression, rather than outcome-driven production. We present \textit{Artly}, a generative AI tool for digital artists. Instead of intervening directly in the act of creation, \textit{Artly} was designed to help artists improve their own work through a combination of personalizable AI feedback and human-authored learning resources. Compared to many productivity-focused contexts, our focus is on art-creation for self-expression that has a stronger requirement for respecting artists' personal style; yet, without appropriate prompting, AI-generated feedback tends to be generic~\cite{loStretchingAIsReach2026,benharrakWriterDefinedAIPersonas2024}. Our goal was therefore to enable users to specify their targeted style without forcing them into effortful and challenging text prompting. Our proposed approach builds on style tags extracted from reference images that users can either upload, search, or generate.
Because image generation is viewed particularly critically by many artists, we investigated it separately by implementing two system variants: one variant with and one without the option to generate variations of the user's own illustration as references. We posed the following two research questions:

\begin{quote}
    \textbf{RQ1:} How do digital artists perceive personalizable AI feedback that is grounded in human-authored learning resources?

    \textbf{RQ2:} How does reference generation affect artists' perceptions and interactions with the tool?
\end{quote}

To address these research questions, we conducted an exploratory between-subjects study with predominantly young but experienced digital artists ($N=38$). Results indicate that, for many participants, \textit{Artly} shifted their perceptions of generative AI, as they experienced AI for the first time not as a potential threat, but as valuable support for artists. Overall, participants perceived our tool as helpful for identifying areas for improvement and for developing new directions for their work. A notable exception concerned the most proficient participants, who reported that the AI feedback lacked sufficient specificity and depth for their level of expertise. The reference-generation feature had less impact than anticipated and appeared to divert attention away from the AI feedback.

In sum, we contribute an exploration of how generative AI can be used for artistic practice in a way that is accepted by artists. Our findings suggest that our feedback-based approach holds promise, while also identifying key limitations and opportunities for improvement.

\section{Related Work}
Integrating generative AI into creative and artistic processes in a way that supports and augments human creativity is a highly active area of research. Often, generative AI is used in creative applications for one-shot generation of polished results~\cite{holznerGenerativeAICreativity2025}, which can induce design fixation around AI outputs~\cite{Wadinambiarachchi_2024_CHI_fixation}, leading to reduced originality and increased homogeneity of final results~\cite{holznerGenerativeAICreativity2025,zhouCreativeScarGenerative2026,andersonHomogenizationEffectsLarge2024}. Moreover, artworks created under heavy AI usage are perceived to lack authenticity~\cite{messerCocreatingArtGenerative2024}. Beyond these effects on creative outputs, one-shot generators also impact the creative process: Artists may become less cognitively engaged~\cite{10.1145/3706598.3713886} and feel disconnected from the process~\cite{oztasReevaluatingCreativeLabor2025} as they perceive a lack of control~\cite{lyuCommunicationHumanAI2022}. 

Among these challenges, the lack of control over image generators has become a major focus of recent work. Apart from technical advances that improve the controllability of image generation models~\cite{Zhang_2023_ICCV_ControlNetLike,Mou_2024_AAAI_T2IAdapter}, various approaches have been proposed in recent years to improve the user interface to these models. The aim is usually to reframe the human-AI interaction from one-shot generation into a co-creation process~\cite{chenInteractiveDrawingAssistant2025,weberDrawinginStepsSupportingCreative2025,chungPromptPaintSteeringTexttoImage2023,dangWorldSmithIterativeExpressive2023,Lawton_2023_IUI_Reframer,Fan_2024_ContextCam}. Some of these proposed interfaces remain closer to the turn-based and prompt-driven interaction paradigm~\cite{chenInteractiveDrawingAssistant2025}, while others re-imagine the interaction more radically~\cite{chungPromptPaintSteeringTexttoImage2023,Lawton_2023_IUI_Reframer}. These co-creation tools increase artists' productivity while giving them considerably more control than common one-shot generation tools. The increased human input may also alleviate many of the other issues mentioned above, but since the production of the artwork is still dominated by AI, some of the challenges may persist, such as the risk of design fixation and homogenization or a perceived lack of authenticity.

Another key thread of research is to use generative AI earlier in the creative process, for ideation rather than for production~\cite{leongParatrouperExploratoryCreation2025,10.1145/3706598.3714148,Mozaffari_2022_GANSpiration,choiCreativeConnectSupportingReference2024}. Common use cases include the generation of new references~\cite{Mozaffari_2022_GANSpiration}, recombination of references~\cite{choiCreativeConnectSupportingReference2024,10.1145/3706598.3714148}, and rapid exploration of the design space~\cite{leongParatrouperExploratoryCreation2025}. These ideation tools are typically considered helpful by participants, but evaluations are mostly based on participants' own perceptions. Dedicated studies on the effect of generative AI on divergent thinking found no positive effect on participants' creativity~\cite{10.1145/3706598.3713886,shinNoEvidenceLLMs2025}, despite participants perceiving AI to be helpful for inspiration and gaining new perspectives~\cite{10.1145/3706598.3713886}. It remains an open question whether user interfaces specifically designed for ideation can increase actual creativity and not only perceived creativity. However, results by both \citet{10.1145/3706598.3713886} and \citet{shinNoEvidenceLLMs2025} suggest that even when generative AI is used for ideation rather than production, there is a risk of cognitive disengagement.

One-shot generation, co-creation, and ideation are the most common creativity use cases for generative AI~\cite{Ko_2023_IUI_artists}. Another conceivable, but far less studied usage of generative AI for creative applications is to provide feedback. Different from the other approaches, AI would not directly intervene into creative processes, but critique users' own work and advise them on how to improve. Users would retain more independence and engagement in their creative process, while AI could potentially still offer meaningful benefits. The feedback approach has been explored in various design applications such as UI~\cite{duanGeneratingAutomaticFeedback2024}, CAD~\cite{longFeedQUACQuickUnobtrusive2026}, and visual design~\cite{liVizCritExploringStrategies2026}. Results suggest that AI-generated feedback can be highly valuable for designers to spot mistakes~\cite{duanGeneratingAutomaticFeedback2024,liVizCritExploringStrategies2026} or boost confidence in their own work~\cite{longFeedQUACQuickUnobtrusive2026,liVizCritExploringStrategies2026}. Beyond design, AI feedback has been explored in domains such as writing~\cite{liuCraftingTextCrafting2026,wettsteinAdaptiveTutoringModalities2026} or programming~\cite{scholzPartneringAIPedagogical2026}, often in educational, but also in productive settings~\cite{benharrakWriterDefinedAIPersonas2024}.

However, AI-generated feedback has not yet been extensively investigated in the context of art-making, even though feedback and critique are common practices in art, both in art education~\cite{barrettStudioCritiquesStudent2000} as well as professional settings~\cite{stockettProfessionalFeedbackArtists2026}. \citet{aritaAssessingLLMsArt2025} have recently shown that LLMs are able to generate feedback that is often hard to distinguish from human-written feedback. AI-generated feedback could address artists' needs for an outside perspective or learning new techniques~\cite{pengExploringOpportunitiesSupport2025}. While related to the design use cases of prior work, the purpose in art-making is rather enjoyment and self-expression than creating a solution to a set of requirements or design principles. The artist's personal style is likely of higher importance in such a setting than in design professions, and any AI-generated feedback needs to respect that personal style rather than being solely based on general notions of a ``good'' solution. 

One example of AI feedback for artists related to our work is ArtKrit by \citet{maComputationalScaffoldingComposition2025}, although their tool has a narrow focus on replicating reference images for improving technical skills rather than creative self-expression. The aim of our work is to explore whether AI feedback can be helpful in such a personal rather than solution-focused setting, with a distinct focus on allowing users to tailor the feedback to their style and needs. Even more closely related to our work is a recent study by \citet{johnstonCoDesigningAIFeedback2026}, who co-designed an AI feedback tool with four visual artists. Similar to our work, their prototype allows participants to adjust different aspects of the feedback. However, beyond simple feedback settings, it only allows users to textually enter their goal for the feedback. We were interested in how users can communicate their targeted art style and tailor the feedback to it beyond using text prompts. A second distinct focus of our work was how to integrate image generation into a feedback tool that is accepted by artists.

\section{System Design and Implementation}
\label{sec:system_design}
We built \textit{Artly} to explore how digital artists would perceive and interact with AI-generated feedback on their own personal work. In this section, we describe our rationale in designing \textit{Artly} (\autoref{sec:design_rationale}), how users would interact with the tool (\autoref{sec:user_journey}), the two variants of it that we built for the study (\autoref{sec:system_variants}), and the technical implementation (\autoref{sec:architecture}).

\subsection{Design Rationale}
\label{sec:design_rationale}
We had two main goals when designing \textit{Artly}: (1) ensuring that users could easily tailor the AI feedback to their personal style and needs, and (2) grounding AI feedback in trustworthy human-authored learning material. 

To be useful, the AI feedback had to be specific to users' personal art style. However, large language model (LLM) outputs can be very generic and unhelpful when not prompted with enough specificity~\cite{benharrakWriterDefinedAIPersonas2024,loStretchingAIsReach2026}. At the same time, artistic concepts may be hard to verbalize. Thus, we sought to support users in communicating their personal style to the tool through style tags extracted from reference images. This approach mirrors traditional art education, where artists develop signature techniques through iterative refinement against established examples. Apart from style, we also aimed to support users in controlling what aspects they wanted to receive feedback on. We did this by presenting AI feedback in a highly structured form and adhering to the ``\textit{overview first, zoom and filter, then details-on-demand}'' mantra~\cite{shneidermanEyesHaveIt1996}, as detailed in \autoref{sec:user_journey}. 

Another key concern with LLMs is the uncertain trustworthiness of their outputs. To help users contextualize the AI-generated feedback, we therefore added educational material created by the first author\footnote{The first author is himself a professional digital artist with over 250k followers on social media channels and teaches digital art.}, including video tutorials on art fundamentals, practical guidelines, illustration process videos, and a brush set. This material was referenced in \textit{Artly}'s deepest feedback level and could also be independently browsed within the interface.

\begin{figure*}[t]
  \centering
  \includegraphics[width=1\textwidth]{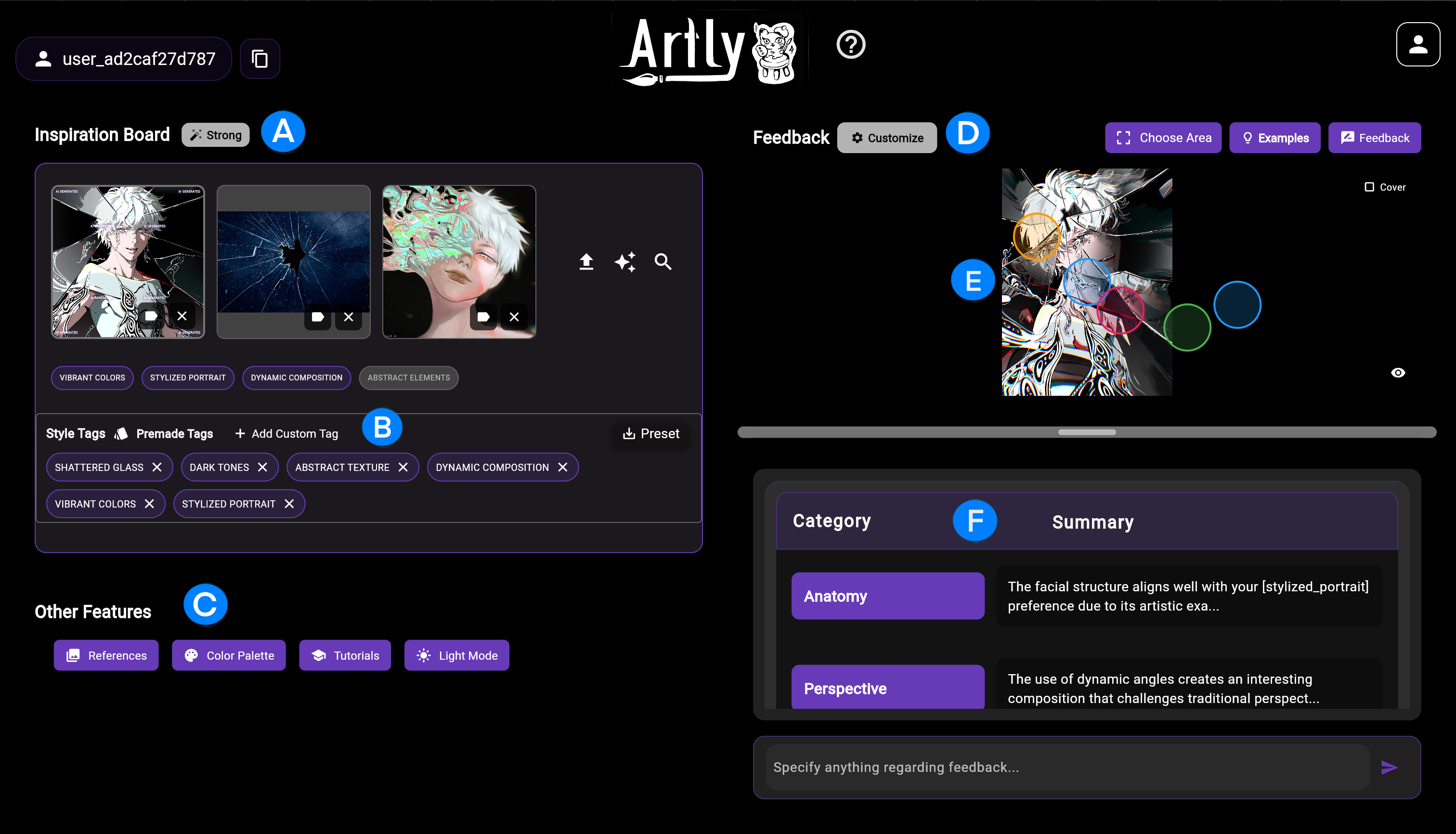}
  \caption{Overview of \textit{Artly}'s main view. The key components of the interface were the inspiration board (A) where users collected references and defined style tags (B), and the feedback area with a summary table (F) that served as entry point for the AI feedback. Users could customize the feedback (D) by (de-)selecting feedback categories, their level of expertise, and whether the feedback should be more creative or technical. The feedback area also displayed the user's own illustration with visual overlays (E) for local feedback. Lastly, the main screen linked to educational material and a color palette generator (C).}
  \Description{A screenshot of the Artly main interface, labeled with letters A through F to indicate key features. The layout is divided into two primary columns. The left column includes the "Inspiration Board" (A) showing a gallery of three reference images with associated tags, the "Style Tags" section (B) containing a list of selectable artistic descriptors like "Shattered Glass" and "Dark Tones," and "Other Features" (C) which provides quick links to references, color palettes, and tutorials. The right column displays the "Feedback" area, featuring a "Customize" button (D) for AI settings, a main image window (E) where the user's illustration is marked with multiple colored circular overlays for localized feedback, and a summary table (F) at the bottom right that provides categorical AI-generated critiques on "Anatomy" and "Perspective." A text input field at the very bottom allows users to specify additional feedback requests.}
  \label{fig:artly_UI}
\end{figure*}
\subsection{System Walkthrough and User Journey}
\label{sec:user_journey}
The main view in \textit{Artly} consisted of three primary sections (\autoref{fig:artly_UI}): an inspiration board, the feedback area, and a collection of non-AI features, including a color palette generator and educational material. The workflow started with users uploading their own illustration (\autoref{fig:teaser}). Users could then collect reference images in the inspiration board, with three options available to them: uploading from their device, retrieval from stock-image-library APIs (Unsplash, Shutterstock, Pexels), and, depending on the system variant (see \autoref{sec:system_variants}), also generating variations of their own artwork using generative AI. Apart from serving as inspiration, the references were also meant to help users communicate their personal art style to \textit{Artly}. For each reference, the tool extracted a set of suggested style tags (e.g., warm tones, realistic anatomy). Users could select and edit these tags to define their target style, ensuring that subsequent AI feedback was aligned with their specific artistic goals.

\begin{figure*}[t]
  \centering
  \includegraphics[trim={0 0.65cm 0 0.7cm},clip,width=0.8\textwidth]{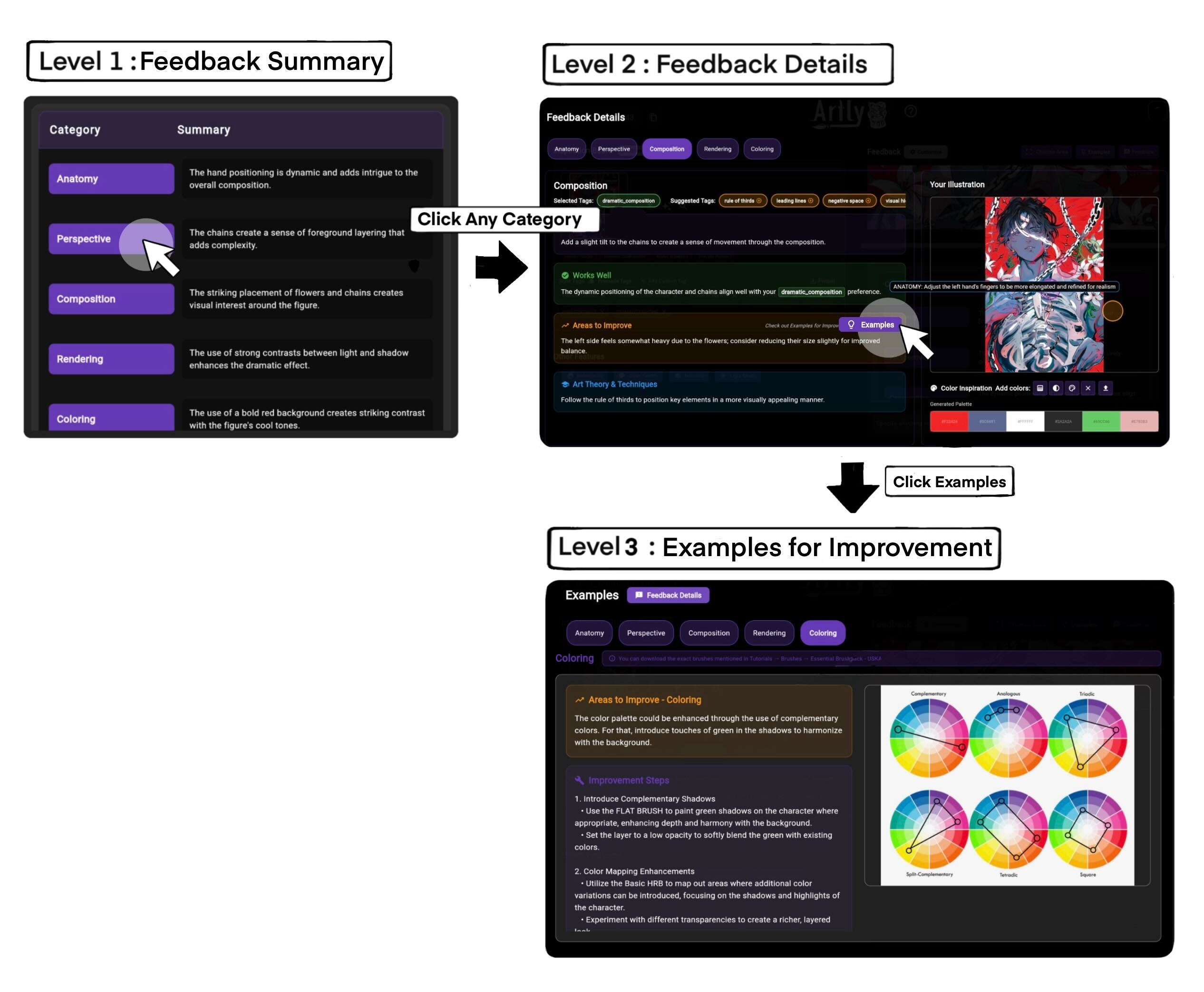}
  \caption{Overview of the three levels of AI feedback in \textit{Artly}. Clicking on one of the feedback categories in the feedback summary (see \autoref{fig:feedback_table} for a larger, readable screenshot) opened the details view for that category (\autoref{fig:feedback_details}). Within the details, clicking on \textit{Examples} in the \textit{Areas to Improve} section opened concrete examples for improvement (\autoref{fig:combined_feedback}).}
  \Description{A flow diagram illustrating the three-level hierarchical structure of AI feedback in the Artly interface. Level 1, "Feedback Summary," shows a high-level table with categories like Anatomy, Perspective, and Composition alongside brief summary text; a cursor is shown selecting the "Perspective" category. An arrow labeled "Click Any Category" leads to Level 2, "Feedback Details." This screen displays granular feedback for a specific category (Composition), organized into "Working Well" and "Areas to Improve" sections. It also features a preview of the user's illustration. A cursor clicks an "Examples" button within the "Areas to Improve" box. A final arrow labeled "Click Examples" leads to Level 3, "Examples for Improvement." This view provides instructional content for a selected category (Coloring), featuring a numbered list of "Improvement Steps" and a visual reference panel containing six different color wheel diagrams representing various color harmony theories.}
  \label{fig:feedback_levels}
\end{figure*}
For the feedback, the system analyzed the user's artwork against the defined style tags and selected categories. The feedback was then presented through a three-level progressive disclosure model (\autoref{fig:feedback_levels}) to allow users to quickly see the type of feedback available to them and explore details on-demand for categories of interest (see \autoref{sec:feedback_examples} for an impression of the concrete feedback contents):
\begin{enumerate}
    \item \textbf{Feedback Summary:} A table displaying a high-level summary of positive aspects across five default categories: anatomy, perspective, composition, rendering, and coloring. Users could select or deselect any category if desired; other available categories included atmosphere, values, textures, and  practical. They could also adjust their level of expertise and whether the feedback should be more creative or technical. This summary table served as entry point for \textit{Artly}'s feedback. 
    \item \textbf{Feedback Details:} More detailed feedback for each category, with sections for creative suggestions (``Creative''), validation of positive aspects (``Works Well''), suggestions for improvement (``Areas to Improve''), and explanations of underlying art concepts (``Art Theory \& Techniques'').
    \item \textbf{Examples for Improvement:} The most granular level, expanding the ``Areas to Improve'' section of the previous level with detailed step-by-step instructions. It translated general feedback (e.g., ``\textit{fix the eye perspective on the face}'') into concrete actions (e.g., ``\textit{use the Liquify tool to adjust the eye line by x amount}''). The steps included references to concrete brushes that were also explained in the human-authored resources. For additional context, the view further referenced educational material from the tutorials.
\end{enumerate}

The educational resources referenced in the third feedback level, such as video guides and brush sets, were also accessible from the main view so that users could explore them independently from the AI feedback. As an additional non-AI tool for further inspiration, the tool also offered a color palette generator that generated several color combinations based on the user's artwork and different color harmonies (analogous, complementary, monochromatic).

Our intended workflow was for users to first collect references, define style tags from the references, and then explore the AI feedback and educational material, as laid out in \autoref{fig:teaser}. However, users were free to use and move back and forth between the tools in any order they wanted. 

\subsection{System Variants}
\label{sec:system_variants}
\begin{figure}[t]
  \centering
  \includegraphics[width=0.5\textwidth]{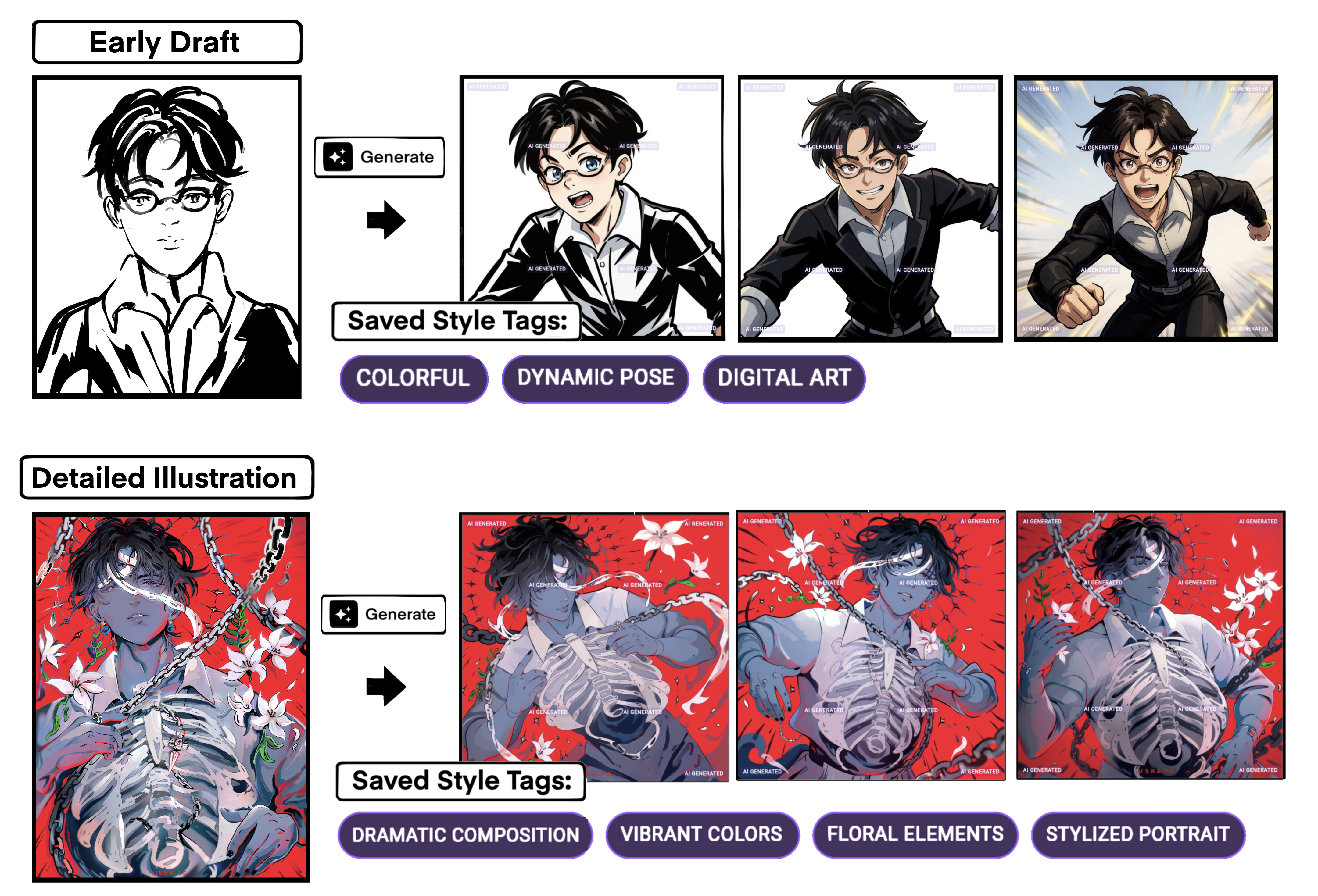}
  \caption{Examples of generated references for an early draft that is still far off from the targeted style, as captured by the saved style tags (top), and generated references for a more worked out illustration (bottom). Generated references were overlaid with watermarks to ensure that users would only use them as references, not as final outputs.}
  \Description{A two-row comparison showing how Artly generates reference images from different stages of an illustration. The top row, labeled "Early Draft," starts with a simple black-and-white line art sketch of a male character with glasses. An arrow pointing to a "Generate" button leads to three color-rendered reference images of the same character in various dynamic poses. Below these generated images are "Saved Style Tags": COLORFUL, DYNAMIC POSE, and DIGITAL ART. The bottom row, labeled "Detailed Illustration," starts with a complex, full-color illustration of a character with a visible ribcage, chains, and lilies. Following a "Generate" button, three new reference images are shown that maintain the core elements and color palette of the original but offer variations in pose and composition. Below these are "Saved Style Tags": DRAMATIC COMPOSITION, VIBRANT COLORS, FLORAL ELEMENTS, and STYLIZED PORTRAIT. All six generated reference images in both rows feature visible watermarks to distinguish them from final user outputs.}
  \label{fig:ref_gen}
\end{figure}
As image generation is seen particularly critically by artists, we wanted to investigate its impact on users' perceptions and interactions separately. We therefore designed two variants of our tool, which we called Weak Mode and Strong Mode. In Weak Mode, \textit{Artly} had no image generation features, and users could only collect references through their own uploads and stock-image-API searches. For the latter, the tool would automatically search for similar images rather then letting users type out search queries. In Strong Mode, users could additionally generate variations of their own illustration as references. The generated references would vary the character's pose and composition and also take users' existing style tags into account (\autoref{fig:ref_gen}). To foster serendipity and to ensure focus on inspiration, users had no further control over the generation process apart from requesting a new generation or editing style tags. They could however, as with uploaded and searched references, extract new style tags from the generated references if they found some new element in the images that they wanted to incorporate into the AI feedback. We suspected that the reference-generation feature would allow users to explore more easily where they wanted to take their artwork next and also communicate these directions more precisely to \textit{Artly} by extracting tags from the generated references, as these would be closer to their own artwork than the non-generated references. At the same time, we expected that the generated references could impair users' sense of agency through design fixation. 

 

\subsection{System Architecture and Prompting}
\label{sec:architecture}
\begin{figure*}[t]
  \centering
  \includegraphics[width=0.75\textwidth]{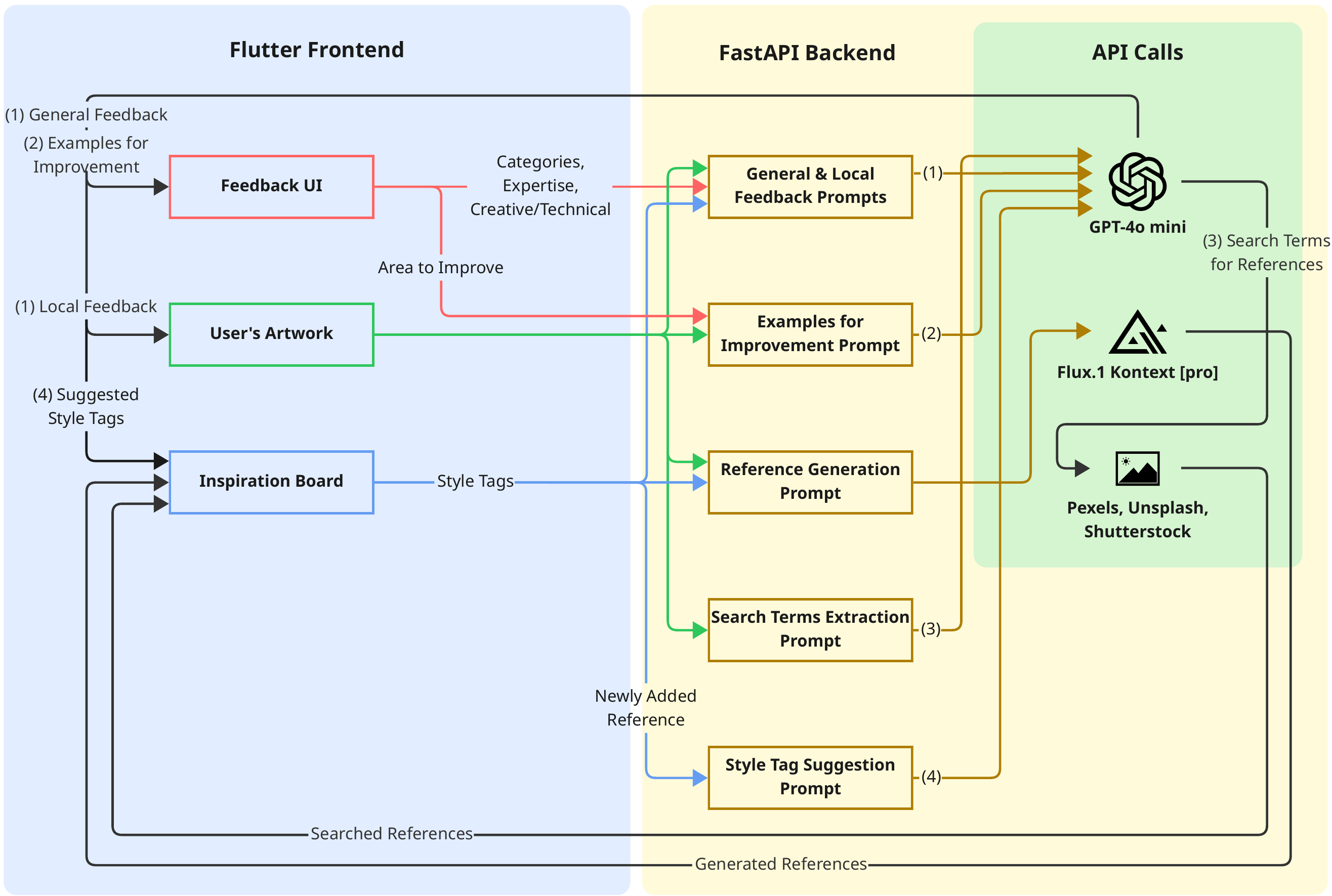}
  \caption{Overview of \textit{Artly}'s architecture and prompting. The arrow labels denote which information is passed between components. Numerical labels provide the mapping between the multiple inputs and outputs of GPT-4o mini. For example: The \textit{Style Tag Suggestion Prompt} receives the \textit{Newly Added Reference} from the \textit{Inspiration Board}, which GPT-4o mini uses to generate \textit{Suggested Style Tags} that are passed back into the \textit{Inspiration Board} for users to review.}
  \Description{An architecture diagram of the Artly system, organized into three vertical columns: Flutter Frontend (left), FastAPI Backend (center), and API Calls (right). The Flutter Frontend includes three main modules: Feedback UI, which sends user categories and expertise levels to the backend. User's Artwork, which sends the user's current image data for processing. Inspiration Board, which sends style tags and newly added references. The FastAPI Backend acts as an intermediary, containing five prompt-handling modules: 'General & Local Feedback Prompts,' 'Examples for Improvement Prompt,' 'Reference Generation Prompt,' 'Search Terms Extraction Prompt,' and 'Style Tag Suggestion Prompt.' These modules receive data from the Frontend and format it for the external APIs. The API Calls column features three service types: GPT-4o mini, which processes prompts to return 'General Feedback,' 'Local Feedback,' and 'Suggested Style Tags' to the frontend, and sends 'Search Terms' to stock image providers. Flux.1 Kontext [pro], which uses generation prompts to return 'Generated References' to the Inspiration Board. Stock providers (Pexels, Unsplash, Shutterstock), which receive search terms from GPT-4o mini to return 'Searched References' to the Inspiration Board. Arrows indicate a continuous loop where user input is processed by the backend and external AI services to provide automated feedback and visual references back to the frontend UI.}
  \label{fig:architecture}
\end{figure*}
The architecture of our prototype consisted of a Flutter frontend, a FastAPI backend that orchestrated prompts and API calls, and a PostgreSQL database to log interaction data. We used GPT-4o mini to generate feedback and Flux.1 Kontext [pro] to generate references. GPT-4o mini was also used to extract suggested style tags from new references added to the inspiration board, as well as search queries from user's artwork for the image-API-based reference-search feature. \autoref{fig:architecture} gives an overview of the role of each API call and which inputs were used for each prompt. We provide the full prompts in \autoref{sec:prompts}.

\section{User Study}
We conducted a between-subjects study to assess participants' perceptions of and interactions with \textit{Artly}’s two variants, assigning participants to either Weak Mode or Strong Mode. Rather than providing illustrations, we asked participants to test \textit{Artly} with their own in-progress illustration. We chose a between-subjects design because a within-subjects design would have required each participant to be working on two illustrations concurrently, or would have introduced carryover effects from reusing the same illustration across both conditions. In the following, we describe our participant recruitment and sample (\autoref{sec:participants}), study procedure (\autoref{sec:procedure}), questionnaire design (\autoref{sec:questionnaire_design}), and data analysis approach (\autoref{sec:data_analysis}).

\subsection{Participants}
\label{sec:participants}
We recruited 41 participants through the first author's personal online network, targeting a pool of young and aspiring artists already engaged with digital art. No interaction logs were recorded for three participants, which we removed from our analysis, leaving 38 participants (18 for Weak Mode, 20 for Strong Mode, see \autoref{sec:participants_details}). The final participants sample was predominantly young ($M = 22.0, SD = 4.62$ years) and experienced digital artists, with 78.9\% reporting three or more years of experience. In terms of self-assessed proficiency, most identified as \textit{Intermediate} (52.6\%) or \textit{Advanced} (23.7\%), with fewer identifying as \textit{Beginners} (15.8\%) or \textit{Professionals} (7.9\%). While most participants (63.2\%) reported prior exposure to AI tools, only 7.9\% used AI tools that were specifically designed for artistic purposes. 57.9\% of participants reported being \textit{concerned} or \textit{very concerned} about the impact of generative AI on the art industry.

\subsection{Procedure}
\label{sec:procedure}
The study was conducted online via a form with a link to the \textit{Artly} web app that was sent directly to participants. Study sessions lasted about 30 minutes and followed a structured workflow:
\begin{enumerate}
\item \textbf{Introduction:} After giving informed consent, participants watched a tutorial video describing the core features of their assigned mode (Weak or Strong).
\item \textbf{Demographics:} Participants completed a brief pre-task questionnaire capturing their artistic background and prior experience with AI-based tools.
\item \textbf{Study task:} Participants used \textit{Artly} with their own original illustrations and progressed through the full system workflow, including gathering references, defining style tags, customizing feedback parameters, receiving feedback, exploring color palettes, and consulting the educational material.
\item \textbf{Evaluation:} Immediately after completing the task, participants completed a post-task questionnaire assessing their experience retrospectively.
\end{enumerate}
The study was approved by \anon[the IRB at an anyonymized institution]{the Ethics Committee of the Faculty of Mathematics, Computer Science and Statistics at LMU Munich}.

\subsection{Questionnaire Design}
\label{sec:questionnaire_design}
The post-task questionnaire consisted of nine five-point Likert-scale items. To assess participants’ perceived control over the interaction, we measured perceived \textit{agency}, perceived \textit{control} over the AI feedback, and perceived \textit{adaptation} to participants' individual style. To capture perceived artistic outcomes, we asked participants to report how \textit{creative} they felt after using the tool, the extent to which it supported the generation of \textit{new ideas and directions}, and its perceived \textit{helpfulness} in improving their illustrations. The two former items were loosely informed by the \textit{Expressiveness} and \textit{Exploration} factors of the Creativity Support Index (CSI)~\cite{cherry2014quantifying}. We did not adopt the CSI verbatim, as it presupposes a tool used to produce an artifact---an assumption that does not align with our feedback-driven approach. Finally, to examine participants’ perceptions of using AI in artistic practice, we asked whether they experienced \textit{discomfort} incorporating AI into their workflow, whether they perceived the \textit{non-AI features as more helpful than the AI features}, and whether using \textit{Artly} influenced their \textit{attitudes toward AI} in the art industry more broadly. In addition to the Likert-scale items, we included three open-ended questions probing (1) why participants’ attitudes toward AI changed or remained unchanged, (2) why they would or would not use \textit{Artly} again, and (3) what they learned from using it. The full questionnaire is provided in \autoref{sec:full_qestionnaire}.


\subsection{Data Analysis}
\label{sec:data_analysis}
Quantitative data included the Likert-scale responses from the questionnaire and data from interaction logs. Several participants used the web app to receive feedback on multiple illustrations during the study. We therefore normalized interaction data by each participant's number of uploaded illustrations to avoid over-representing participants with repeated usage. Given our small sample size and the exploratory nature of our analysis, we refrained from inferential statistics and only report descriptive statistics including effect sizes. We report medians ($Md$) and interquartile ranges ($IQR$) for central tendency and spread. For comparisons between two groups, we report Cliff's Delta ($\delta$) as effect size, and for comparisons between more than two groups, we report Epsilon-squared ($\epsilon^2$). For additional transparency, we show individual data points in Figures~\ref{fig:likert_expertise}--\ref{fig:likert_gen}.

We analyzed qualitative data from the open-ended questionnaire items using thematic analysis \cite{blandfordAnalysingData2016}. Initially, one author performed open coding on all written responses, resulting in a codebook of 31 codes. To ensure analytical rigor, a second author independently applied it to all written responses. We evaluated inter-rater reliability on the overlapping text segments coded by both coders, with a Cohen’s Kappa of 0.575, indicating moderate to substantial agreement \cite{landisMeasurementObserverAgreement1977}. After this independent coding, the two coders met to discuss all discrepancies, refining code definitions and reaching full consensus. Finally, the first coder iteratively grouped the codes into three themes. 
Where appropriate, we report counts of participants who made a certain statement in our description of the themes.

\section{Findings}
We report our findings along two threads, reflecting our research questions: participants' perceptions of \textit{Artly} (\autoref{sec:perceptions}), and the effects of the reference-generation feature on participants' interactions and perceptions (\autoref{sec:gen_effects}). We also present a small qualitative sample of how participants incorporated the AI feedback into their artwork (\autoref{sec:follow-up}).

\subsection{Participant Perceptions}
\label{sec:perceptions}
We identified three themes in our thematic analysis of participants' free text answers: (1) \textit{Artly} reframed AI from perceived threat to valuable support for artists (\autoref{sec:theme_reframing}). (2) The main value of our tool was in the easy access to new ideas and areas for improvement (\autoref{sec:theme_easy_access}). (3) Different from other participants, the most advanced participants perceived little value in using \textit{Artly} (\autoref{sec:theme_experts}). In the following, `W-$x$' denotes a participant from the Weak Mode group. In the Strong Mode group, some participants used the reference-generation feature, while others did not (more details in \autoref{sec:gen_effects}). `S-R-$x$' denotes a participant who used reference generation, and `S-NR-$x$' a participant who did not utilize that feature. Quotes from participants' responses are slightly edited in terms of spelling correction to improve readability.

\subsubsection{Reframing AI from threat to support}
\label{sec:theme_reframing}
We explicitly asked in one of the open-ended questions whether participants' attitude toward AI in the art industry had changed after using \textit{Artly}. Eleven participants mentioned in their answers that tools like ours do not address their fundamental concerns about AI, including copyright infringement, replacement of artists, and, less frequently mentioned, environmental impact. In contrast, 20 participants described how \textit{Artly} prompted them to reconsider their attitude towards AI, offering them ``\textit{a new picture towards AI where it supports rather than replaces artists}''~(S-R-8), and where ``\textit{AI is used as a tool, rather than replacing the entire creative process}''~(S-R-6), ``\textit{mak[ing] it feel like it was working with you, not against}''~(W-1). Three participants discussed both sides in their response, displaying a nuanced perspective:
\begin{quote}
    ``\textit{My attitude has changed somewhat. Using Artly showed me that AI can be a valuable supporting tool when used responsibly, especially for feedback and improvement without replacing creativity. However, I still have concerns about the broader impact of AI in the art field.}''~(W-16)
\end{quote}

Of those participants mentioning a positive attitude change, three participants described how they already had an abstract idea for how AI could benefit artists, and that our tool was a concrete illustration of that idea:
\begin{quote}
    ``\textit{I always kind of imagined it could be used as an aid rather than a replacement. Getting to use it though really solidified it.}''~(S-R-5)
    
    ``\textit{I believe there are ways to use AI as actually a helpful tool. However that just applies to Artly (so far and from what I know)}''~(W-10)
\end{quote}
However, for most participants, \textit{Artly} caused a shift in perspective, as the possibility of using AI as supportive tool for art was entirely new to them:
\begin{quote}
    ``\textit{I always thought it would just steal, not serve or benefit artists in any way.}''~(W-3)

    ``\textit{With how negatively AI is viewed specially in the art world it would be a lie to say I wasn't nervous when I originally looked into this. [...] My attitude definitely changed, as this is an AI tool intended to help artists better work on their passions without worry of losing their individuality or style.}''~(S-NR-7)
\end{quote}
The next theme describes how concretely the tool supported participants.


\subsubsection{Easy access to new ideas and areas for improvement}
\label{sec:theme_easy_access}
The AI feedback feature was predominantly well-received, with 25 participants explicitly commenting on its helpfulness. Participants particularly valued the low barriers to access, as AI feedback is readily available with a certain level of competence (``\textit{I want people to criticize my art but most people I know don't understand art theory}''~(W8)) and is perceived as non-judgmental (``\textit{I felt like I was learning how to improve my art rather than feel belittled for trying to learn how to improve.}''~(W-1)). The combination with educational resources further increased the perceived helpfulness of the AI feedback and was explicitly praised by three participants. Compared to traditional resources, \textit{Artly}'s resources felt more tailored to users' specific needs:
\begin{quote}
    ``\textit{Everytime I try to find YouTube tutorials or artists' help online, it never quite feels like it's helping with exactly what I need [...]. Artly really points out the exact areas I can improve and how. I'm really excited to get back into drawing now :))}''~(S-R-5)
\end{quote}
Our tool delivers a wide range of feedback. While this could be overwhelming in principle, participants’ responses did not suggest overload; instead, they described selectively focusing on feedback aligned with their current needs:
\begin{quote}
    ``\textit{I'm not too worried about the color portion yet, however, through the anatomy portion of the feedback, I learned a lot and will definitely be using it and applying it to my art. (Color will be kept in mind and obviously I will use that, too, just not as much as the anatomy, not yet, haha!)}''~(W-1)
\end{quote}

From the feedback, participants mainly gained new ideas and directions for their art, as well as increased awareness of mistakes and areas for improvement: 
\begin{quote}
    ``\textit{When i am feeling stuck and need inspiration to improve my art; to get new ideas in where to go with my project.}''~(W-13)
    
    ``\textit{It helps you see the artwork from a different perspective leading you to find any potential flaws, especially since after staring for hours and hours at the same piece one can't really see the mistakes and needs a fresh look to find them.}''~(S-NR-4)
\end{quote}
As a result, participants reported feeling more creative, broadening of their artistic perspective, and sharpening of their critical assessment skills:
\begin{quote}
    ``\textit{Considering the fact that I personally often use only muted colors, it explained and showed me how incorporating vibrant color accents could benefit my own style as well. Even though I'm not a `bright and vibrant' colors guy, I'll definitely give it a try!}''~(S-NR-4)
    
    ``\textit{Receiving direct feedback has made me sharpen my own critical thinking of looking at art.}''~(W-5)
\end{quote}
However, not all participants perceived these features of the tool as benefits, with one participant stating that ``\textit{I still feel AI will only funnel my choice into one direction and not expand my horizon that much.}''~(S-NR-8). We investigated this discrepancy further in the following theme.


\begin{figure*}[t]
    \centering
    \includegraphics[width=0.66\linewidth]{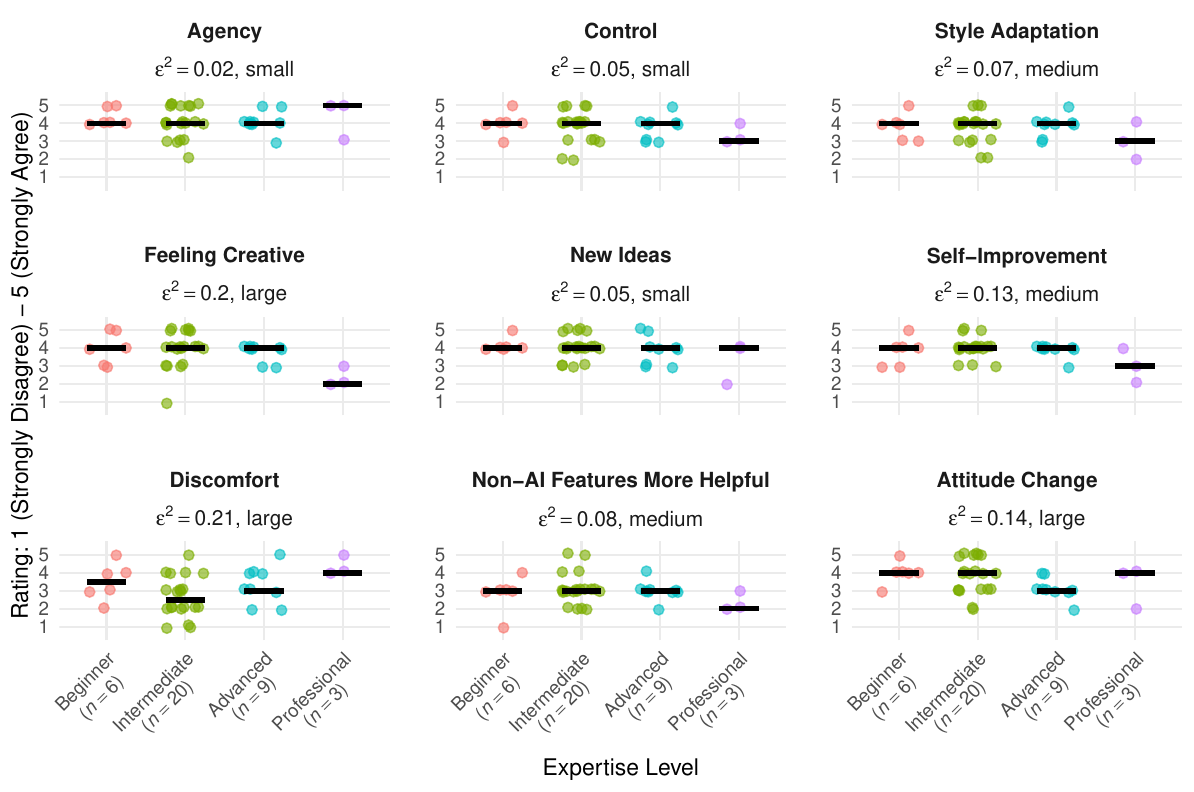}
    \caption{Self-reports of participants by level of art expertise. Horizontal lines indicate medians. Effect size interpretations follow the thresholds proposed by \citet{fieldDiscoveringStatisticsUsing2013}.}
    \label{fig:likert_expertise}
    \Description{A grid of nine scatter plots displaying participant self-reports by four art expertise levels: Beginner (n=6), Intermediate (n=20), Advanced (n=9), and Professional (n=3). The vertical axis for all charts represents a rating from 1 (Strongly Disagree) to 5 (Strongly Agree). Each plot includes individual data points, a horizontal bar for the median, the epsilon squared value, and an effect size interpretation. 1. Agency (epsilon squared 0.02, small effect): medians are 4 for most, 5 for professionals. 2. Control (epsilon squared 0.05, small effect): medians are 4 for most, 3 for professionals. 3. Style Adaptation (epsilon squared 0.07, medium effect): medians are 4 for most, 3 for professionals. 4. Feeling Creative (epsilon squared 0.20, large effect): medians are 4 for most, dropping to 2 for professionals. 5. New Ideas (epsilon squared 0.05, small effect): medians are consistently high at 4. 6. Self-Improvement (epsilon squared 0.13, medium effect): medians are 4 for most, 3 for professionals. 7. Discomfort (epsilon squared 0.21, large effect): medians range from 2.5 for intermediate to 4 for professionals. 8. Non-AI Features More Helpful (epsilon squared 0.08, medium effect): medians are 3 for most, 2 for professionals. 9. Attitude Change (epsilon squared 0.14, large effect): medians are 4 for most, 3 for advanced.}
\end{figure*}
\subsubsection{Diminishing returns for most advanced participants}
\label{sec:theme_experts}
While positive comments about \textit{Artly} predominated in participants' responses, a few participants found it not helpful at all, stating that the feedback was too vague, generic, or fundamental:
\begin{quote}
    ``\textit{It gives me the fundamentals, which I, in that case, don't need personally. [...] I also believe the explanations are often a bit vague and lack context.}''~(W-10)

    ``\textit{It didn't give me much on what I should change while I see the artwork needs some improvements.}''~(S-NR-1)

    ``\textit{Advice was a bit too generic, and actually not describing what was in the picture. Not very targeted.}''~(S-NR-8)
\end{quote}
However, several participants stated the exact opposite, stating that the AI feedback was ``\textit{specific}''~(W-11, S-NR-9) and ``\textit{insightful}''~(S-NR-7, S-R-6). To better understand this discrepancy, we examined participants' self-reported levels of artistic expertise and found that all three participants who explicitly found the AI feedback unhelpful self-identified as Professional (W-10, S-NR-8) or Advanced (S-NR-1). In contrast, all participants who described the AI feedback as being specific or insightful self-identified as Beginner or Intermediate artist in the demographics section of the study. Two quotes shed further light on this expertise divide. Participant W-10 (Professional) remarked that ``\textit{there is just no way of telling if the AI understands what your vision really is.}'', whereas Participant S-NR-7 (Intermediate) stated:
\begin{quote}
    ``\textit{Some feedback I found very insightful was some of the recommendations or tips pertaining to make my gradient smoother as well as adding more `pop' to the eye focal points I have on my piece. I understand a little more how to elevate my art piece without changing it entirely.}''~(S-NR-7)
\end{quote}
Apparently, more advanced artists tended to have highly specific intentions. Feedback that was not fully aligned with those intentions was considered unhelpful. Less advanced artists tended to be less specific, making it easier for AI to add to their process.

Motivated by this expertise divide in our qualitative data, we examined participants' Likert-scale responses stratified by their art expertise (\autoref{fig:likert_expertise}), confirming that expertise strongly influenced participants' perceptions in our sample. In particular, on many items, Beginner, Intermediate, and Advanced participants rated similarly, while Professional participants' ratings deviated from the other expertise levels. They reported having less control over the feedback and that it showed poorer adaptation to their style, suggesting that they found it harder to steer the AI towards feedback that met their high expectations. Nevertheless, compared to other participants, they tended to rate the AI features as more helpful relative to \textit{Artly}'s non-AI features. We suppose that to Professional participants, the educational resources might have been even more basic than the AI feedback. Consequently, Professional participants also tended to feel less creative after using our tool and tended to perceive less improvement to their artworks from using it. 

Discomfort over using AI and attitude change were two items that deviated from the pattern of Professional participants expressing different opinions than the other expertise levels. Both items displayed a U-shape where Intermediate and Advanced participants had a tendency for lower ratings than Beginner and Professional participants. While effect sizes for both items were large, these patterns were less directly reflected in the free text responses and thus harder to interpret.

Overall, both the Likert-scale and the free text responses consistently indicate that \textit{Artly} was helpful for most participants, but was not able to meet the high demands of the most advanced participants.

\subsection{Effects of Reference-Generation Functionality}
\label{sec:gen_effects}
\begin{figure*}[t]
    \centering
    \includegraphics[trim={0 1cm 0 0},clip,width=0.68\linewidth]{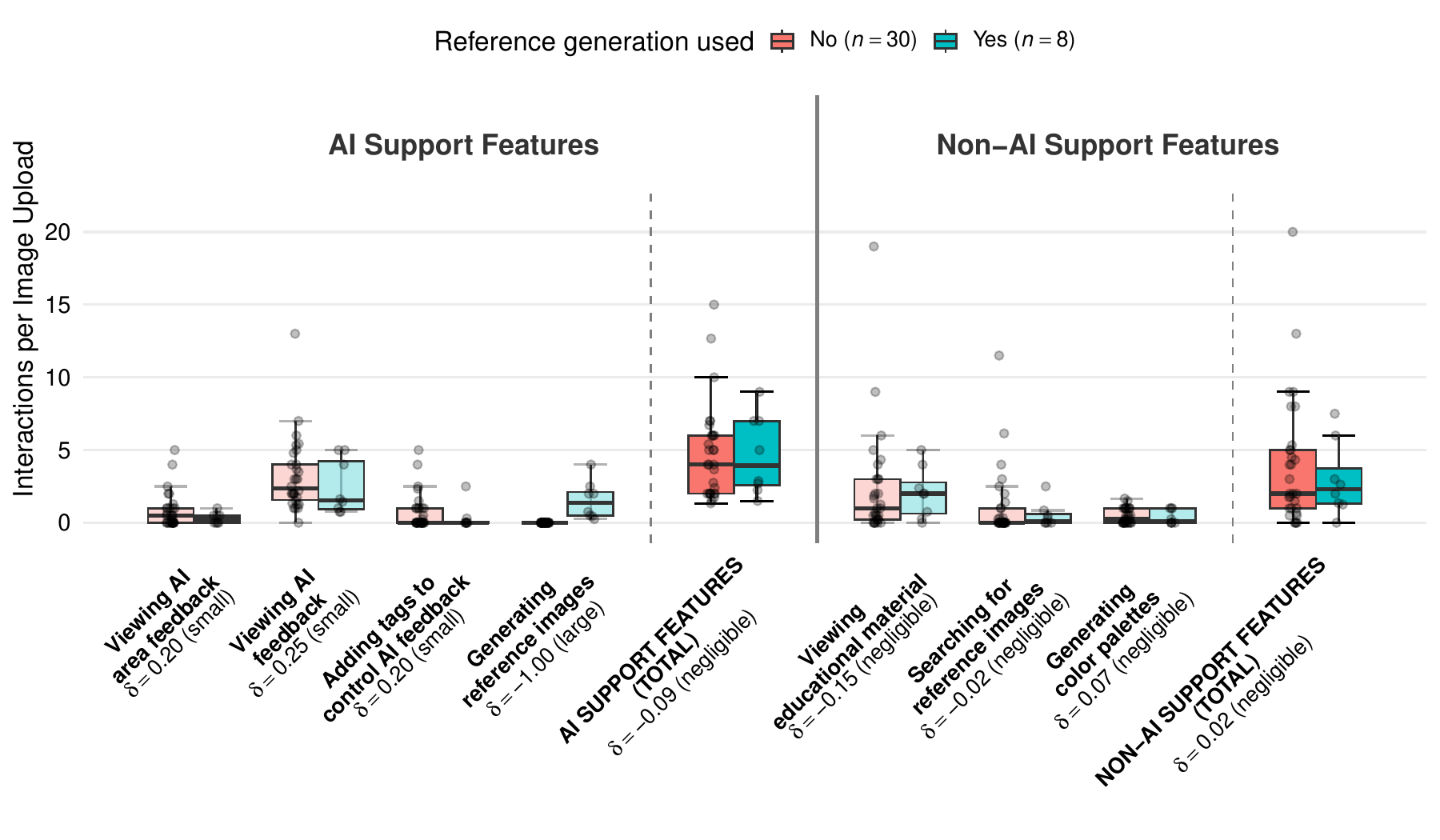}
    \caption{Interactions with AI and non-AI features of participants who used/did not use reference generation. Lighter colors indicate subcategories of the overarching \textit{AI} and \textit{Non-AI Support Features} categories. Gray dots represent individual data points. Effect size interpretations follow the thresholds proposed by \citet{romano2006appropriate}. $\delta<0$ indicates higher values for \textit{Yes}, $\delta>0$ higher values for \textit{No}.}
    \label{fig:interaction}
    \Description{A box plot chart comparing the frequency of interactions per image upload between two participant groups: those who used the reference-generation feature (n=8, indicated in teal) and those who did not (n=30, indicated in pink). The vertical axis represents the number of interactions per image upload, ranging from 0 to 20. The horizontal axis is split into two sections: AI Support Features and Non-AI Support Features. Under AI Support Features, subcategories include: Viewing AI area feedback (delta 0.20, small effect), Viewing AI feedback (delta 0.25, small effect), Adding tags to control AI feedback (delta 0.20, small effect), and Generating reference images (delta negative 1.00, large effect). The total interactions for AI Support Features show a negligible difference with a delta of negative 0.09. Under Non-AI Support Features, subcategories include: Viewing educational material (delta negative 0.15, negligible), Searching for reference images (delta negative 0.02, negligible), and Generating color palettes (delta 0.07, negligible). The total for Non-AI Support Features also shows a negligible difference with a delta of 0.02. Individual data points are plotted as gray dots over the box plots. Delta values less than zero indicate higher values for the group that used reference generation, while values greater than zero indicate higher values for those who did not.}
\end{figure*}

\begin{figure*}[t]
    \centering
    \includegraphics[width=0.66\linewidth]{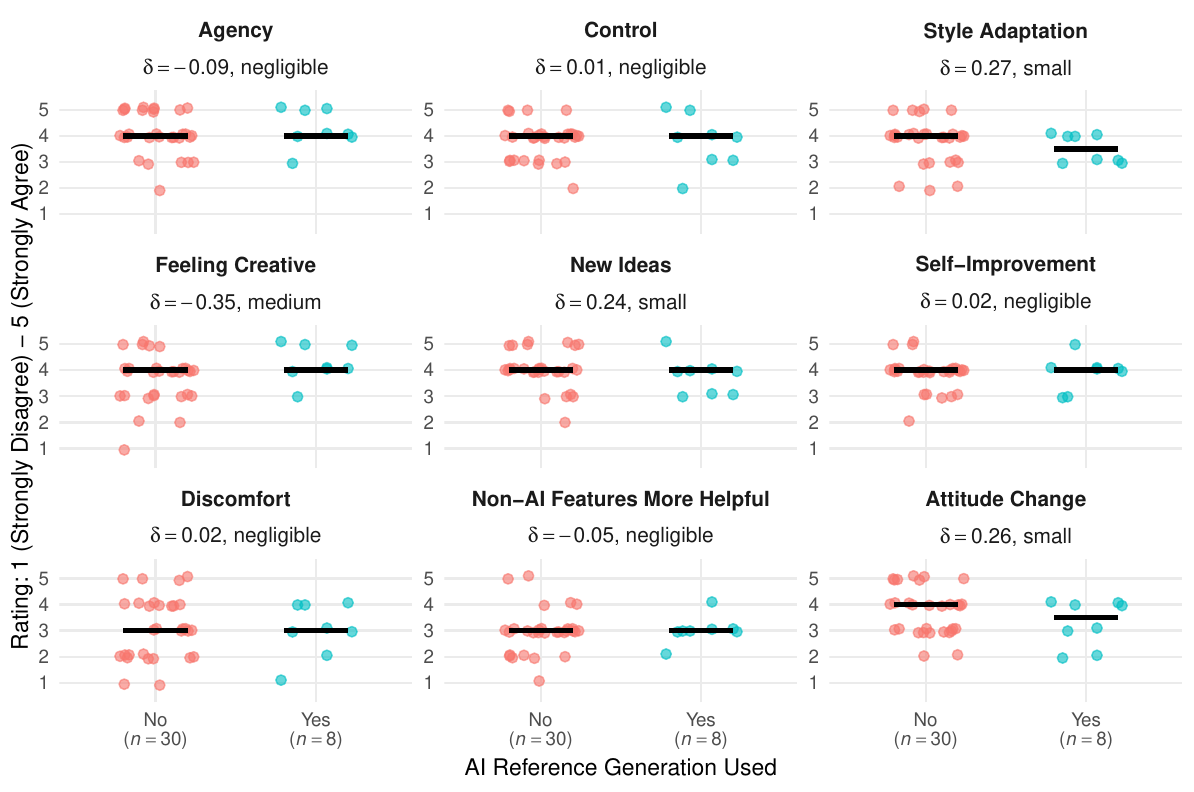}
    \caption{Self-reports of participants who used/did not use reference generation. Horizontal lines indicate medians. Effect size interpretations follow the thresholds proposed by \citet{romano2006appropriate}. $\delta<0$ indicates higher values for \textit{Yes}, $\delta>0$ indicates higher values for \textit{No}.}
    \label{fig:likert_gen}
    \Description{A grid of nine scatter plots comparing self-reported ratings between two groups: participants who did not use AI reference generation (n equals 30, shown in pink) and those who did (n equals 8, shown in teal). The vertical axis for all charts represents a rating from 1 (Strongly Disagree) to 5 (Strongly Agree), and black horizontal bars indicate the median for each group. The delta values indicate the effect size, where values less than zero favor the "Yes" group and values greater than zero favor the "No" group. 1. Agency: Delta is negative 0.09, negligible effect. Both groups have a median of 4. 2. Control: Delta is 0.01, negligible effect. Both groups have a median of 4. 3. Style Adaptation: Delta is 0.27, small effect. The "No" group median is 4, while the "Yes" group is 3.5. 4. Feeling Creative: Delta is negative 0.35, medium effect. Both groups have a median of 4. 5. New Ideas: Delta is 0.24, small effect. Both groups have a median of 4. 6. Self-Improvement: Delta is 0.02, negligible effect. Both groups have a median of 4. 7. Discomfort: Delta is 0.02, negligible effect. Both groups have a median of 3. 8. Non-AI Features More Helpful: Delta is negative 0.05, negligible effect. Both groups have a median of 3. 9. Attitude Change: Delta is 0.26, small effect. The "No" group median is 4, and the "Yes" group is 3.5.}
\end{figure*}

We were interested in how image generation would affect participants' interactions and their perceptions of the tool. However, upon inspecting the interaction logs, we noticed that only eight out of 20 Strong Mode participants used the reference-generation feature, i.e., the other 12 participants used \textit{Artly} the same way as participants who used the Weak Mode. Since we were interested in the effects of \textit{using} the reference-generation feature rather than the effects of its mere presence\footnote{One could argue that the mere presence of the feature may have had an effect despite disuse; for instance, it may have evoked negative sentiments. However, qualitative responses provided no relevant indications.}, we decided to analyze by whether participants used the feature or not rather than by study condition, which is also reflected in Figures~\ref{fig:interaction} and \ref{fig:likert_gen}.

First, we explored whether usage of the reference-generation feature changed how often other features of the tool were used, in particular whether a shift between AI- and non-AI-support features would occur (\autoref{fig:interaction}). The interaction logs revealed only negligible differences in the total usage of AI and non-AI features. However, a small shift of usage \textit{within} the AI features occurred, as participants who used the reference generation engaged less frequently with the AI feedback: They viewed both the area-specific feedback ($Md\ (IQR)$: 0.25 (0.00-0.50) vs. 0.50 (0.00-1.00)) and the overall feedback ($Md\ (IQR)$: 1.54 (0.94-4.25) vs. 2.35 (1.56-4.00)) less frequently\footnote{\autoref{fig:interaction} also shows that participants who used reference generation added fewer tags to control the feedback. However, due to a bug in our logging, \textit{Adding tags to control AI feedback} only includes custom-defined tags and no tags added through reference images, leading to an incomplete comparison.}. Hence, reference generation did not divert attention away from non-AI features like human-authored educational material, but from the other AI features.

Next, we examined how the usage of the reference-generation feature affected participants' Likert-scale responses (\autoref{fig:likert_gen}). Again, many items only had negligible differences. However, participants who used the generation feature showed a slight trend toward reporting that \textit{Artly} adapted to their style more poorly, which is consistent with the insight that they tended to use fewer tags to control the AI feedback. As for how creative participants felt after using \textit{Artly} and how much it helped them to find new ideas and directions, the median score was $Md=4$ for both items and both groups. However, effect sizes indicate trends of opposing directions: Participants who used reference generation were more likely to feel more creative, but less likely to gain new ideas after using our tool. 
Lastly, participants who used reference generation were also slightly less likely to report a change in attitude towards AI. This is again consistent with their slightly lower engagement with the AI feedback, which appeared to be the main driver for participants' positive sentiments (see \autoref{sec:perceptions}).

\subsection{Artwork Outcomes}
\label{sec:follow-up}
\begin{figure}[t]
    \centering
    \includegraphics[width=0.47\textwidth]{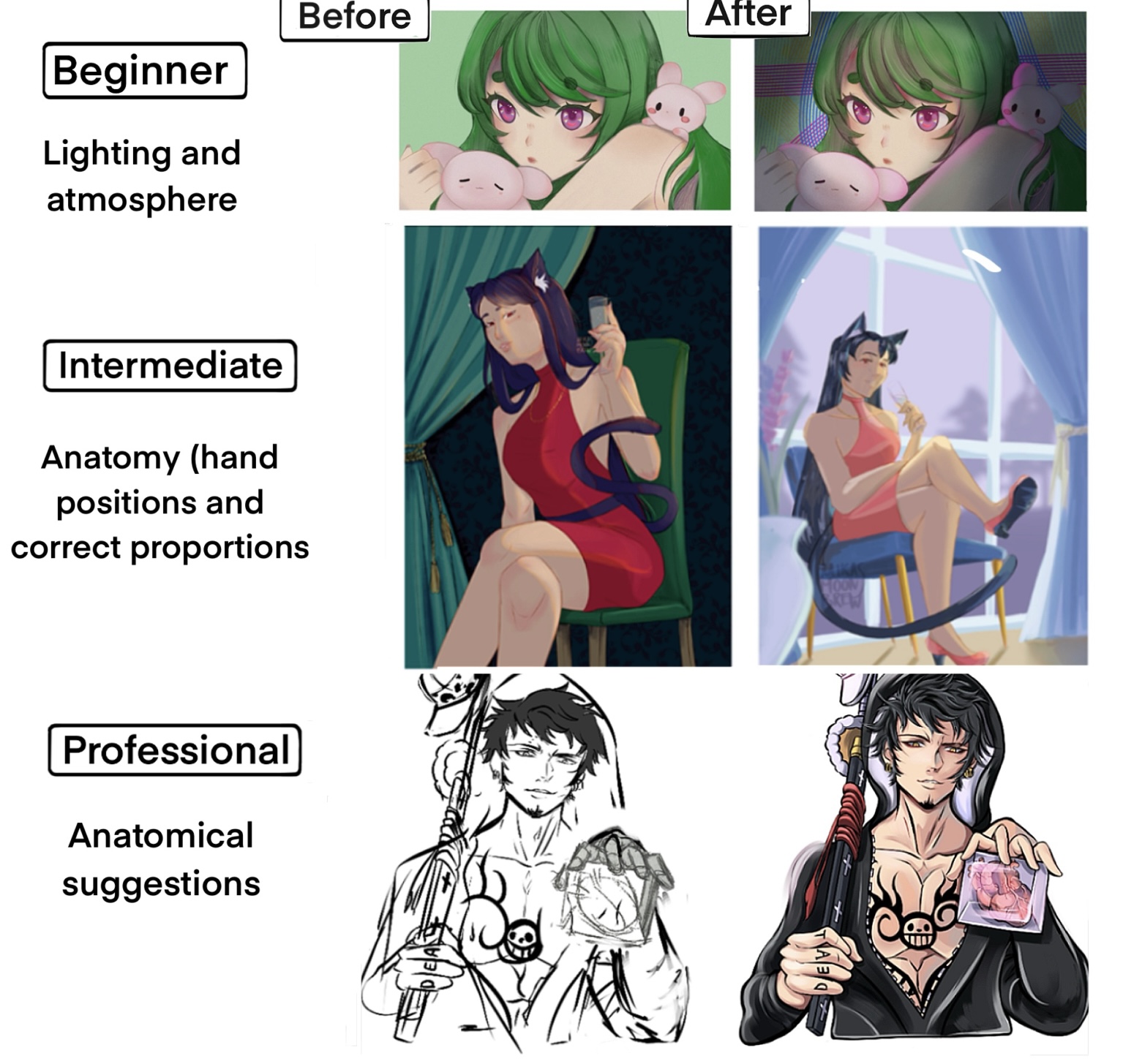}
    \caption{Participants' illustrations before and after using \textit{Artly}. For each pair, participants' self-rated expertise is annotated, as well their answers to the question ``\textit{What is one thing that Artly suggested you that helped you the most?}'' \copyright~Participants, used with permission.}
    \label{fig:follow-up}
    \Description{A comparative figure titled Before and After, showing four examples of character illustrations before and after receiving suggestions from Artly. Each example includes the artist's self-rated expertise and the specific suggestion they found most helpful. Example 1: A Beginner artist focused on Lighting and atmosphere. The before image displays an anime-style character with flat green lighting, while the after image shows the same character with vibrant purple and pink ambient lighting and deeper shadows. Example 2: An Intermediate artist focused on Anatomy, specifically hand positions and correct proportions. The before image shows a character in a red dress with a rigid pose in a dark room. The after image shows the character in a more natural, relaxed pose sitting by a window with significantly improved lighting. Example 3: A Beginner artist focused on Proportions adjustment on a simple sketch. The before image is a rough black and white line drawing of a face, and the after image shows the same sketch with more balanced and symmetrical facial features. Example 4: A Professional artist focused on Anatomical suggestions. The before image is a high-quality ink sketch of a character. The after image is a fully rendered, colored, and professional-grade digital painting of that character.}
\end{figure}
During off-boarding, we asked participants to optionally send us their artwork before and after iterating on it using \textit{Artly}'s feedback; four participants followed this request, three of which gave explicit permission to include their artwork in a scientific publication (\autoref{fig:follow-up}). The illustrations demonstrate a wide range of ways in which participants integrated the tool into their workflows: Some participants used it to receive feedback on early drafts, while others sought feedback on how to refine more final artworks. Further, some participants directly implemented its suggestions, whereas in other follow-up illustrations, \textit{Artly}'s suggestions were less immediately visible as they only made up a small part of the overall changes. Without more explicit context, it is challenging to infer whether the additional changes were also inspired by interacting with \textit{Artly}, or results of participants' independent ideas. Still, the cases demonstrate that participants thought beyond what was directly suggested by the AI feedback.

\section{Discussion}
Our results show that participants' overall sentiments of \textit{Artly} were positive, with most participants perceiving it as helpful for learning and improving their work without limiting their agency or overriding their personal style. Tools like \textit{Artly} may not address artists' concerns around copyright infringements or job displacement, but for many participants, our tool demonstrated for the first time that generative AI can be used for artists' benefit. This highlights that not only model training, but also deployment currently largely fails to meet artists' demands. Our results suggest that the feedback approach is one promising direction to further explore the design of generative-AI tools that are accepted by artists. However, further work is required to assess whether AI feedback tools like ours objectively help artists to refine their skills and artworks, as our study relied on participants self-ratings, which may not always reflect objective performance~\cite{pretzSelfperceptionsCreativityNot2014,kaufmanSelfassessmentsCreativityNot2019}. Longitudinal studies would also be of importance in this regard.

A particularly salient result from our study is the divide between the most proficient participants and other proficiency levels in our sample. This result contributes to the common debate in both artistic domains and beyond about whether generative AI is more beneficial or harmful for novices or experts. Several studies find that novices experience larger gains in productivity and outcome quality when using generative AI than experts~\cite{noyExperimentalEvidenceProductivity2023,houDoubleEdgedRolesGenerative2025,chenLargeLanguageModel2024}, yet it seems that in many cases, experts' existing skills still help them produce better results with AI than novices~\cite{ZhouLee_2024_PNASnexus,shinNoEvidenceLLMs2025}. Novices are further thought to be more exposed to negative implications such as skill degradation or job displacement~\cite{liUserExperienceDesign2024}. These findings typically pertain to generative-AI applications where AI is used to produce artifacts. Our results suggest that shifting the focus from production to feedback might benefit a surprisingly wide range of participants, from beginners to fairly advanced artists. It may also help in both earlier as well as later stages of the creative process, whereas generative AI is typically found to be more useful for earlier stages like ideation~\cite{houDoubleEdgedRolesGenerative2025,tsaoPerceptionsIntegrationGenerative2025}. 

However, for the most advanced participants in our sample, \textit{Artly} was not helpful. Based on our qualitative findings, we assume that this is not only a matter of model performance, where today's models are not capable enough to provide artistic insights at human-expert level. Instead, our findings suggest that a major question is how these most proficient users can more effectively communicate their highly specific intents to the AI. The challenge is that expert knowledge is often tacit~\cite{Polanyi1966} and hard to externalize. We attempted to address this by allowing users to express their style non-verbally through references, but our style tags approach was apparently not expressive enough to capture the nuance that experts expect. Future work should explore interaction patterns that are more effective for expert users. This may require systems that can observe users' own actions more granularly~\cite{sonDemystifyingTacitKnowledge2024}. 
 
One of our objectives for this study was to explore how image generation can be integrated into the feedback approach, which we operationalized through the reference-generation feature. Overall, reference generation did not affect usage and perceptions much. To begin with, it was used and discussed less frequently than we anticipated, probably because it was not an essential component of \textit{Artly}'s workflow. However, given that image generation is often seen particularly critically by artists~\cite{kawakamiImpactGenerativeAI2024,penaRejectionResentmentGenerative2025}, it can be viewed as positive that reference generation did not affect participants' sense of agency, control, or discomfort, suggesting that reference generation is a way of using image generation that is acceptable to artists. For those participants who did use reference generation, it appeared to divert attention away from the AI-feedback features, despite our attempt to integrate both features by allowing users to extract style tags from AI-generated references. The most notable outcome of this shift in usage was a moderate positive effect on participants' feeling of creativity, but curiously, also a small negative effect on the perceived amount of new ideas. One potential explanation is that generated references functioned as prompts for creative experimentation, while feedback tended to provide concrete, technically actionable suggestions that translated into new directions for participants' artworks. Future work should explore how both aspects can be combined rather than traded off against each other. For instance, an alternative way to combine image generation with feedback could be to visualize the textual feedback on demand. 

Overall, our findings suggest that using generative AI for feedback rather than production is a design pattern well suited to domains like art, where the act of creation itself holds intrinsic value. This could also translate to other creative domains like writing, where users may want to retain autonomy and ownership, but AI support often diminishes it~\cite{draxlerAIGhostwriterEffect2024,kreminskiDearthAuthorAISupported2024,qinTimingMattersHow2025}. For instance, \citet{rezaCoWritingAIHuman2025} found that feedback may fit writer needs well in certain contexts, but that it is an under-represented design strategy in AI writing tools. Given that generic, unhelpful AI feedback has also been observed in writing~\cite{benharrakWriterDefinedAIPersonas2024,loStretchingAIsReach2026}, we assume that personalizing feedback to users' specific context would also be the main challenge in other domains beyond art, especially for expert users. This will likely remain true even with more capable models, since the usefulness to experts depends on how well they can externalize their highly specific intents, which are internal to the user, not to the model.

We caution that our quantitative results should be interpreted with care due to the small and unbalanced sample size. Moreover, participants were recruited from the first author's social media followers, which may have introduced a hard-to-quantify bias in favor of the generally positive sentiment toward \textit{Artly}. That said, our sampling strategy allowed for high ecological validity, as all our participants had already been engaged with digital art and brought their own illustrations to the study, rather than working on an artificial study task.

\section{Conclusion}
We set out to explore how generative AI can be used to support artistic purposes rather than outcome-driven creative work. With \textit{Artly}, we explored a feedback-driven approach that is more restrained than many recent generative-AI creativity-support tools. Our results indicate that personalizable AI feedback grounded in human-authored learning material is a promising way to offer significant value to artists in terms of learning and new directions without taking away the actual drawing work. Given the positive reception of our tool by participants, we suggest that more research in this direction is warranted to better address the concerns and needs of groups that so far perceive generative AI mostly as threat, such as digital artists, but possibly also beyond. Directions highlighted by our results include more effective use of image-generation features that complement rather than compete with feedback functionalities, better support for highly proficient users to externalize their specific intents, and deeper understanding of objective versus perceived learning and creativity outcomes.

\section*{Generative AI Usage Disclosure}
Gemini was used to generate the R code to produce Figures \ref{fig:likert_expertise}--\ref{fig:likert_gen}. Analytical and data visualization decisions were made by the authors, while the implementation in R was done by iteratively prompting Gemini according to the authors' decisions.

\bibliographystyle{ACM-Reference-Format}
\bibliography{bibliography,references_tony}

\appendix

\section{Prompts}
\label{sec:prompts}
This section provides all generative-AI prompts used in \textit{Artly}. Sections~\ref{sec:prompt_general}--\ref{sec:prompt_search} were used to prompt GPT-4o mini, \autoref{sec:prompt_generation} to prompt FLUX.1 Kontext [pro]. The inputs to each prompt are detailed in \autoref{fig:architecture}.

\subsection{General Feedback}
\label{sec:prompt_general}
You are an expert art mentor analyzing this specific artwork. If the user has style preferences, only category-relevant tags are used when providing feedback for each specific category. Otherwise, feedback is provided solely based on what is observed in the artwork. Inputs: LEVEL INSTRUCTION: \textit{[levelInstruction]} DETAIL INSTRUCTION: \textit{[detailInstruction]} \textit{[categoriesSection]}

Analyze this artwork and provide feedback according to the level and detail settings above. Be precise about what is visible in the image.

CRITICAL: Use exactly the specified output format. Do not use markdown, headers, or special formatting. Do not add any text beyond what is defined by the format.

CRITICAL: The feedback style, complexity, and level of detail must strictly follow the LEVEL and DETAIL instructions. For style matching, only reference tags that are relevant to the specific category being analyzed. Always reference concrete visual details from the image. Do not use markdown symbols or section headers.

CRITICAL TAG FORMATTING: Whenever tags are mentioned in the feedback, they must be written in the form \texttt{[tag\_name]}. Plain text tag lists such as ``style tags: tag1, tag2'' must never be used. Phrases such as ``fits well with your style tags'' are prohibited. Instead, formulations such as ``aligns with your [specific\_tag] preference'' must be used.

CRITICAL AREAS TO IMPROVE FORMATTING: For all \emph{Areas to Improve} entries, the following structure must be used exactly:

-- One sentence describing the overall improvement area. For that, specific implementation instructions with intuitive directions such as ``a little bit'', ``slightly'', or similarly simple guidance.

Examples include improving color harmony through analogous schemes, enhancing depth using atmospheric perspective, or refining facial proportions through minor adjustments.

MANDATORY: In Long detail mode, all lines defined by the output format must be included, including the \emph{Style Match} line for each category. No lines may be skipped, and no additional text may be added.

CRITICAL FOR CATEGORY-SPECIFIC STYLE TAGS:
Only tags that are relevant to the category currently being analyzed may be mentioned. In positive feedback sections, only category-relevant tags may be referenced, and the same restriction applies to improvement suggestions. All tag references must use the \texttt{[tag\_name]} format. Tags from unrelated categories must never be mentioned. If no relevant tags can be identified, no tags should be mentioned at all.

CRITICAL FOR STYLE MATCH:
When relevant category tags are available, the format must be: \emph{Style Match: Selected Tags: [tag\_name] -- specific comparison analysis}.  
When no relevant tags are available, the format must be exactly: \emph{Style Match: Selected Tags: None. Suggested tags: [tag1, tag2, tag3]}.  
The phrase ``Selected Tags: None'' must always be used consistently and must not be repeated elsewhere. Suggested tags must be short descriptors consisting of one or two words. Already selected tags must never be suggested again, and tags from other categories must not be referenced.

SPECIAL INSTRUCTION FOR THE COLORING CATEGORY:
If the artwork appears to be black and white, monochromatic, or line art without color, the feedback must explicitly recommend exploring palette generation by stating: ``check out Color Inspiration for palette generation'' or a closely matching formulation that includes the term ``Color Inspiration''.

\subsection{Local Feedback}
\label{sec:prompt_highlight}
Analyze this artwork and identify 4--6 specific areas that need attention. For each area:
1. Write a PRECISE, ACTIONABLE description with exact numbers AND qualitative guidance  
2. Choose ONE of these types: anatomy, coloring, composition, perspective  
3. Give coordinates as percentages (0.0 to 1.0) for:  
\begin{itemize}
    \item x: horizontal position (0.0 = left, 1.0 = right)
    \item y: vertical position (0.0 = top, 1.0 = bottom)
    \item width: area width (0.1 to 0.3)
    \item height: area height (0.1 to 0.3)
\end{itemize}

CRITICAL GUIDELINES:
\begin{itemize}
    \item COMBINE exact numbers with qualitative guidance (e.g., ``Add warm orange tones and adjust more saturation by in Hue setting for better harmony'')
    \item Include specific colors, directions, or measurements PLUS ``adjust by a bit'' type suggestions
    \item Focus on CONCRETE changes with both precision and flexibility
    \item Try to cover different aspects of the artwork (don't focus on just one type)
    \item Ensure areas are well-spaced and DO NOT overlap at all
    \item Keep descriptions concise but specific (15--25 words max)
    \item Avoid placing circles too close to image edges (keep x,y between 0.1 and 0.9)
    \item Prioritize the most important aspects that would help improve the artwork
    \item If an area has no meaningful feedback, DO NOT include it
\end{itemize}

EXAMPLES OF GOOD FEEDBACK:
\begin{itemize}
    \item ``Make the eyes a little bit bigger for better proportion''
    \item ``Move the left shoulder slightly to the right for more natural positioning''
    \item ``Add some warmer orange tones to enhance the mood''
    \item ``Make the shadows a bit darker for more depth''
    \item ``Soften the edges slightly for better balance''
\end{itemize}

Return ONLY a JSON array with this exact format:  
\texttt{[{"description": "specific actionable feedback", "type": "anatomy", "x": 0.3, "y": 0.2, "width": 0.2, "height": 0.2}]}

\subsection{Examples for Improvement}
\label{sec:prompt_examples}
USING THE SPECIFIC AREAS TO IMPROVE BELOW, GIVE ME AN EXACT METHOD ON HOW TO IMPROVE THESE ISSUES USING EXACT DIGITAL ART TOOLS IN PROCREATE.

FOCUS CATEGORY: \$category

SPECIFIC AREAS TO IMPROVE FOR \$category:  
\$areasToImprove

CRITICAL: ALL 5 STEPS MUST BE ABOUT \$category IMPROVEMENT ONLY. Do not mention or include steps for any other categories like anatomy, perspective, composition, rendering, or coloring unless the focus category is specifically one of those.

AVAILABLE BRUSHES TO USE:

\begin{enumerate}
    \item Outline Brush -- Creates outline in the edges when drawing.  
    Good for: Long Hair Outline, Quick Silhouettes outline for smaller details, accessories, objects, etc.  
    Tip: Don't Overlay -- Erase, or draw with one stroke.

    \item Main Sketch -- Good for: Early Sketches, Good Stabilisation, Line Variability, Early Flat Coloring, Adding different Values.

    \item Sharp Details -- Good for adding sharp details, small lines like for eyes, hair strands, edges.  
    Tip: Use as Eraser to sharpen or detail your sketch lines.

    \item Basic HRB -- Good for: Soft edges, gradients, smudges paint look.  
    Tip: Combine soft and sharp edges by drawing and erasing with this brush (Good for Flames, soft color transitions).

    \item FLAT BRUSH -- Textured Hard Square brush.  
    Good for: color mapping, rough details, adding different transparencies of the color.

    \item AIRBRUSH -- Good for: Gradients, soft transitions, glow.  
    Tip: Use Lasso selections for precise outcome.

    \item TEXTURES -- Collection of favorite texture brushes.  
    Good for final rendering, adding diversity in textures, adding painterly strokes, glow, watercolor-like effect, smudging.
\end{enumerate}

Please provide a structured improvement guide in this exact format (do NOT include any category header):

\begin{enumerate}
    \item \texttt{[Step Title]}
    \begin{itemize}
        \item \texttt{[Specific action with exact Procreate setting names from the list above]}
        \item \texttt{[Additional detail or setting]}
    \end{itemize}

    \item \texttt{[Step Title]}
    \begin{itemize}
        \item \texttt{[Specific action with exact Procreate setting names from the list above]}
        \item \texttt{[Additional detail or setting]}
    \end{itemize}

    \item \texttt{[Step Title]}
    \begin{itemize}
        \item \texttt{[Specific action with exact Procreate setting names or brush names from the list above]}
        \item \texttt{[Additional detail or setting]}
    \end{itemize}

    \item \texttt{[Step Title]}
    \begin{itemize}
        \item \texttt{[Specific action with exact brush names from the list above]}
        \item \texttt{[Additional detail or setting]}
    \end{itemize}

    \item \texttt{[Mention Useful Procreate tools that you can experiment with for this category (example: for Colors: mention Curves, and Saturation, for Anatomy: Liquify, for Composition: Grid, Liquify, etc)]}
\end{enumerate}

CRITICAL INSTRUCTIONS:
\begin{itemize}
    \item Each step must DIRECTLY ADDRESS one of the specific improvement areas mentioned above
    \item Focus ONLY on fixing the exact issues listed in ``SPECIFIC AREAS TO IMPROVE FOR \$category''
    \item ALL 5 STEPS MUST BE ABOUT \$category ONLY -- do not switch to other categories
    \item Use ONLY the exact brush names from the list above (Outline Brush, Main Sketch, Sharp Details, Basic HRB, FLAT BRUSH, AIRBRUSH, TEXTURES)
    \item Make each step actionable and specific to addressing those \$category improvement points
    \item Mention layer modes and settings that help fix the specific \$category issues
    \item Do not include generic advice -- target the exact \$category problems identified
    \item FORBIDDEN: Do not mention steps for anatomy, perspective, composition, rendering, or coloring unless \$category is specifically one of those categories
\end{itemize}

IMPORTANT: Do not include the text ``Specific action:'', ``Additional detail:'', or any asterisks in your response. Just provide the information directly in bullet points. Each step should clearly state which improvement area it addresses.

FINAL REMINDER: This is about \$category improvement ONLY. Every single step must be related to improving \$category specifically. Do not deviate to other art categories.

\subsection{Style Tag Suggestion}
\label{sec:tag_suggestion}
Analyze this artwork and extract 6 specific visual tags that describe what you actually observe in the image.

Look at these specific aspects:
- VISUAL STYLE: What artistic approach is used? (realistic, stylized, anime, cartoon, painterly, sketch-like, digital, traditional, etc.)
- COLOR CHARACTERISTICS: What stands out about the colors? (vibrant, muted, warm, cool, monochromatic, high-contrast, pastel, etc.)
- COMPOSITIONAL ELEMENTS: What catches the eye? (detailed, minimalist, dynamic, balanced, dramatic, soft, sharp, etc.)
- SUBJECT/CONTENT: What is depicted? (portrait, landscape, character, abstract, fantasy, nature, urban, etc.)

Return ONLY 4 specific tags as a comma-separated list. Each tag should be 1-2 words, descriptive of what you actually see in this specific image, not generic style categories.

Focus on observable visual characteristics that would help someone find similar artwork.

Examples of GOOD specific tags: ``dramatic\_lighting'', ``watercolor\_texture'', ``character\_portrait'', ``warm\_tones'', ``detailed\_background'', ``soft\_shading'', ``fantasy\_creature'', ``urban\_scene''

Examples of BAD generic tags: ``digital\_art'', ``modern'', ``creative'', ``realistic''

\subsection{Reference Search Term Extraction}
\label{sec:prompt_search}
Describe this image in 1--3 words that would be most effective for finding similar reference images.  
Focus only on the main subject and style. Be concise and specific.  
Example: ``portrait sketch monochrome'' or ``anime girl colorful''

\subsection{Reference Generation}
\label{sec:prompt_generation}
Create a new pose variation of this character while maintaining their identity. Style tags: \{style\_tags\_str\}. Keep the same character design and features, create a completely different pose and composition, maintain the same art style and quality, ensure the character is recognizable but in a new position, add dynamic elements to make the pose more interesting

\section{Examples of AI Feedback}
\label{sec:feedback_examples}
\begin{figure*}[t]
  \centering
  \includegraphics[width=0.5\textwidth]{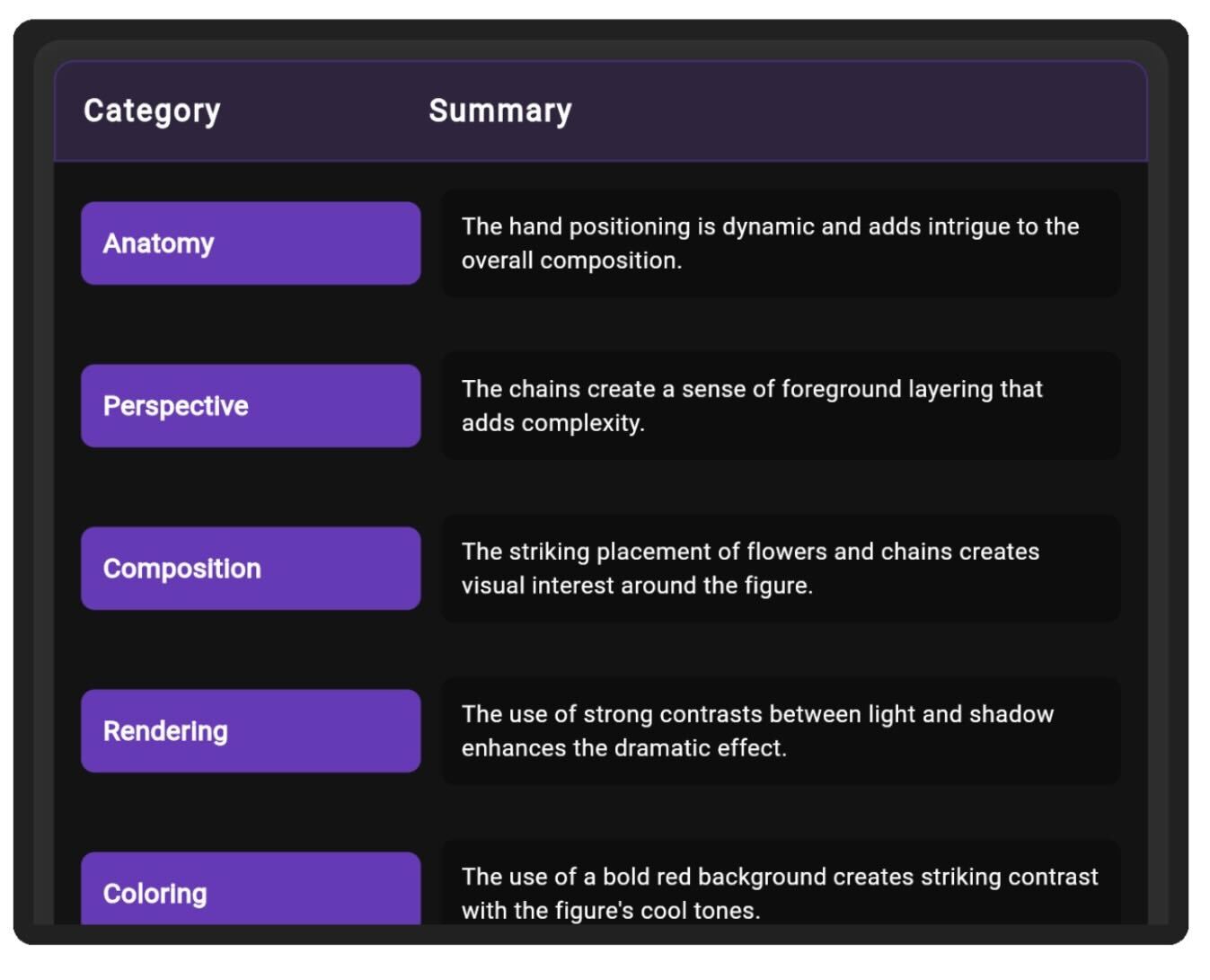}
  \caption{Feedback summary table. This is the entry point for the AI feedback. The short summaries are generated with the general feedback prompt given in \autoref{sec:prompt_general} by extracting the first sentence of the output for each feedback category.}
  \Description{A screenshot showing the Feedback summary table, which serves as the entry point for AI feedback. The table consists of two columns: Category and Summary. It contains five entries. The first entry is Anatomy, with a summary noting that the hand positioning is dynamic and adds intrigue to the overall composition. The second entry is Perspective, with a summary stating that chains create a sense of foreground layering that adds complexity. The third entry is Composition, with a summary highlighting that the striking placement of flowers and chains creates visual interest around the figure. The fourth entry is Rendering, with a summary stating that the use of strong contrasts between light and shadow enhances the dramatic effect. The fifth entry is Coloring, with a summary explaining that the use of a bold red background creates a striking contrast with the figure's cool tones.}
  \label{fig:feedback_table}
\end{figure*}

\begin{figure*}[t]
  \centering
  \includegraphics[width=\textwidth]{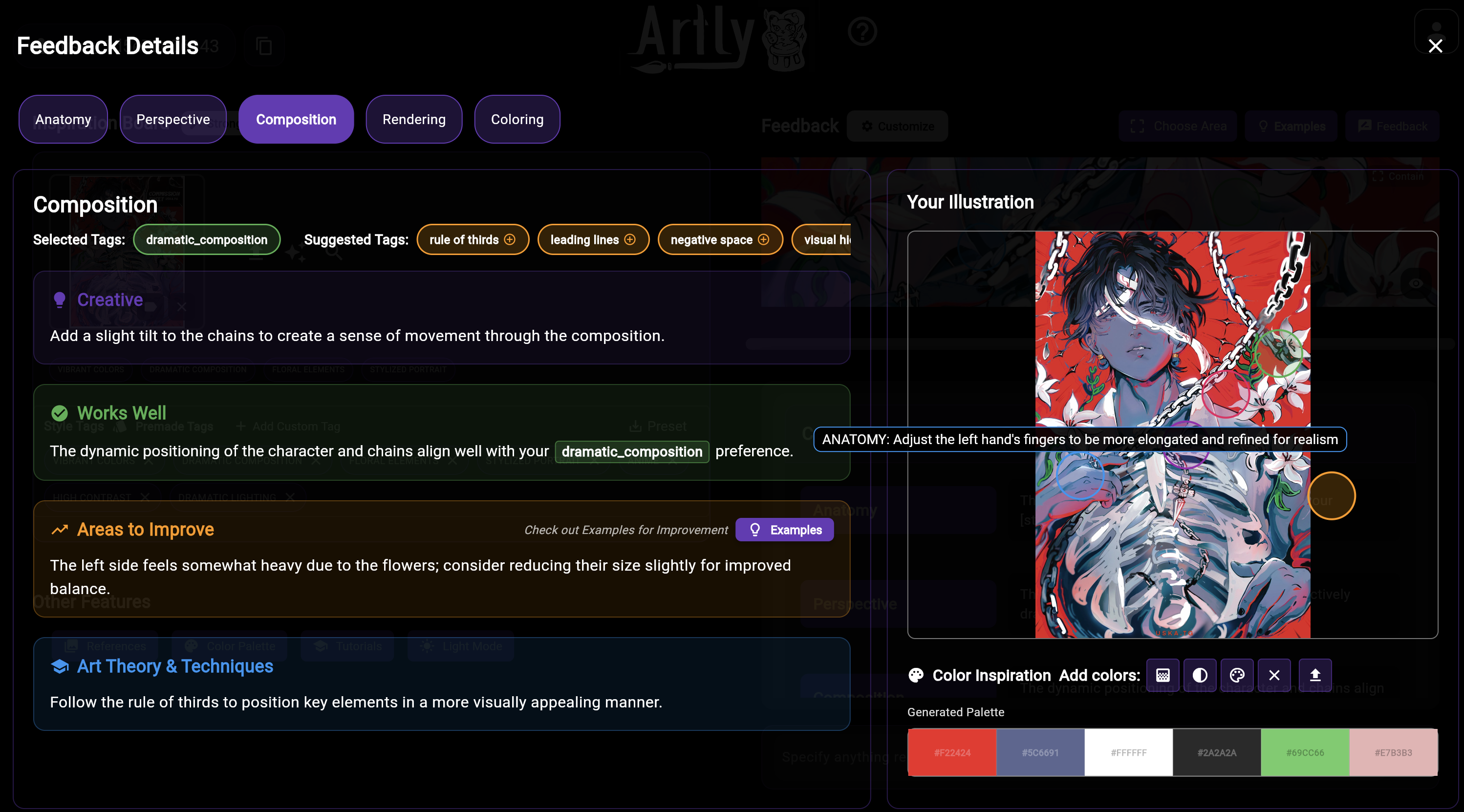}
  \caption{Feedback details view. This view expands the feedback from the summary table. The more detailed feedback is generated with the general feedback prompt given in \autoref{sec:prompt_general}. The overlay feedback on the illustration is generated with with the local feedback prompt given in \autoref{sec:prompt_highlight}.}
  \Description{A screenshot of the Feedback Details view in the Artly interface, showing detailed feedback for the selected Composition category. At the top, five category tabs are displayed: Anatomy, Perspective, Composition, Rendering, and Coloring, with Composition highlighted. The left side of the screen contains specific feedback sections: Selected Tags showing dramatic composition, and Suggested Tags including rule of thirds, leading lines, negative space, and visual balance. Below the tags, four feedback boxes are shown: 1. Creative: suggests adding a slight tilt to the chains for movement. 2. Works Well: notes the dynamic positioning aligns with the selected style. 3. Areas to Improve: mentions the left side feels heavy due to flowers and suggests reducing their size. 4. Art Theory and Techniques: provides advice on using the rule of thirds. The right side features the user's illustration with a specific local feedback bubble stating: Anatomy, Adjust the left hand's fingers to be more elongated and refined for realism. Below the illustration is a Color Inspiration section with a generated palette of six color swatches: red, gray-blue, white, dark gray, green, and light pink.}
  \label{fig:feedback_details}
\end{figure*}

\begin{figure*}[t]
  \centering
  \begin{subfigure}{\textwidth}
    \centering
    \includegraphics[width=0.93\linewidth]{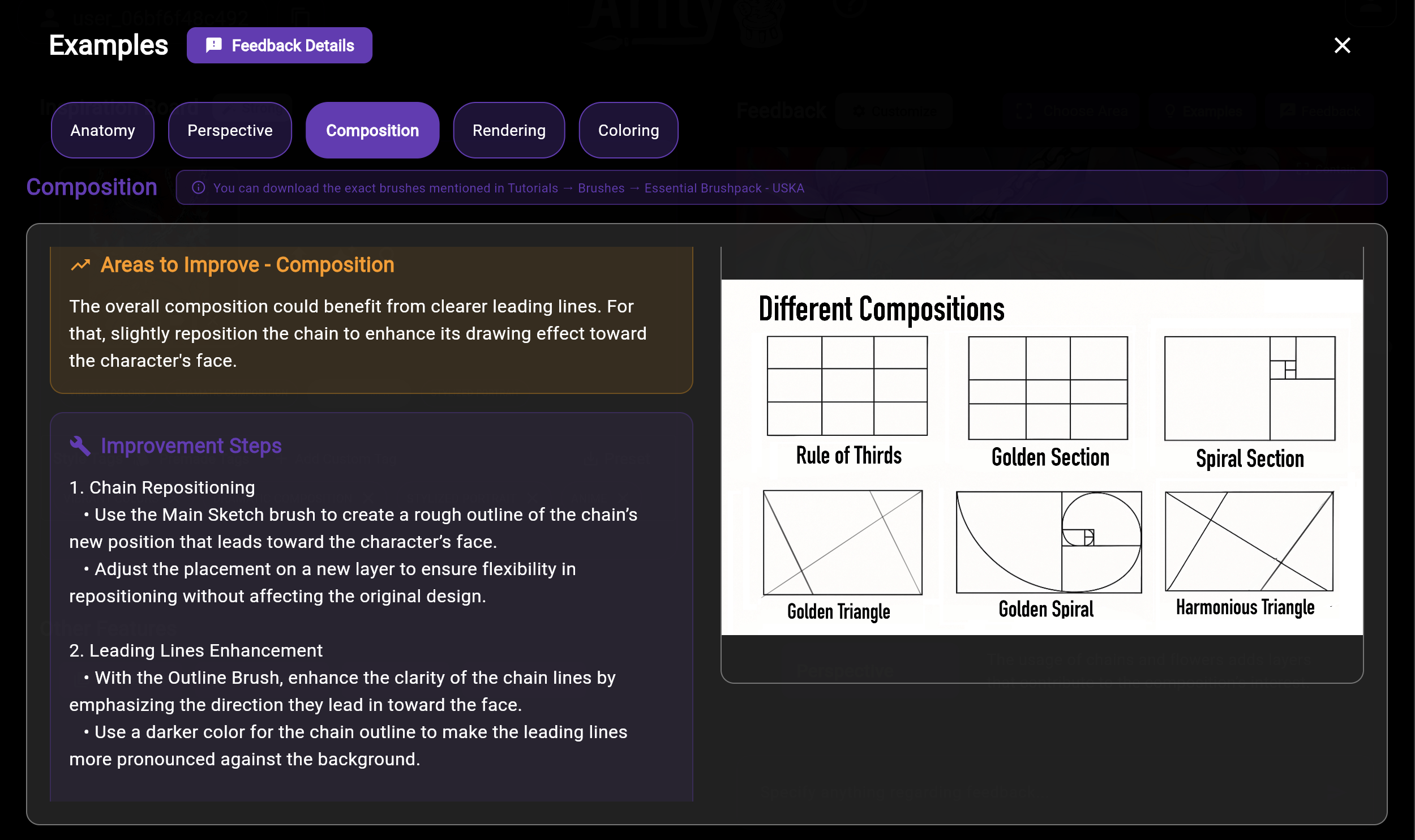}
    \caption{Composition improvements.}
    \label{fig:feedback_composition}
    \Description{A screenshot of the Examples view in the Artly interface for the Composition category. The left side of the screen contains text-based guidance under two headers: Areas to Improve - Composition and Improvement Steps. The text suggests repositioning chain elements to act as leading lines toward the character's face and provides a two-step technical process using specific digital brushes. The right side features a visual reference panel titled Different Compositions, which displays six labeled diagrams of compositional theories: Rule of Thirds, Golden Section, Spiral Section, Golden Triangle, Golden Spiral, and Harmonious Triangle.}
  \end{subfigure}
  
  \vspace{0.3cm} 

  \begin{subfigure}{\textwidth}
    \centering
    \includegraphics[width=0.93\linewidth]{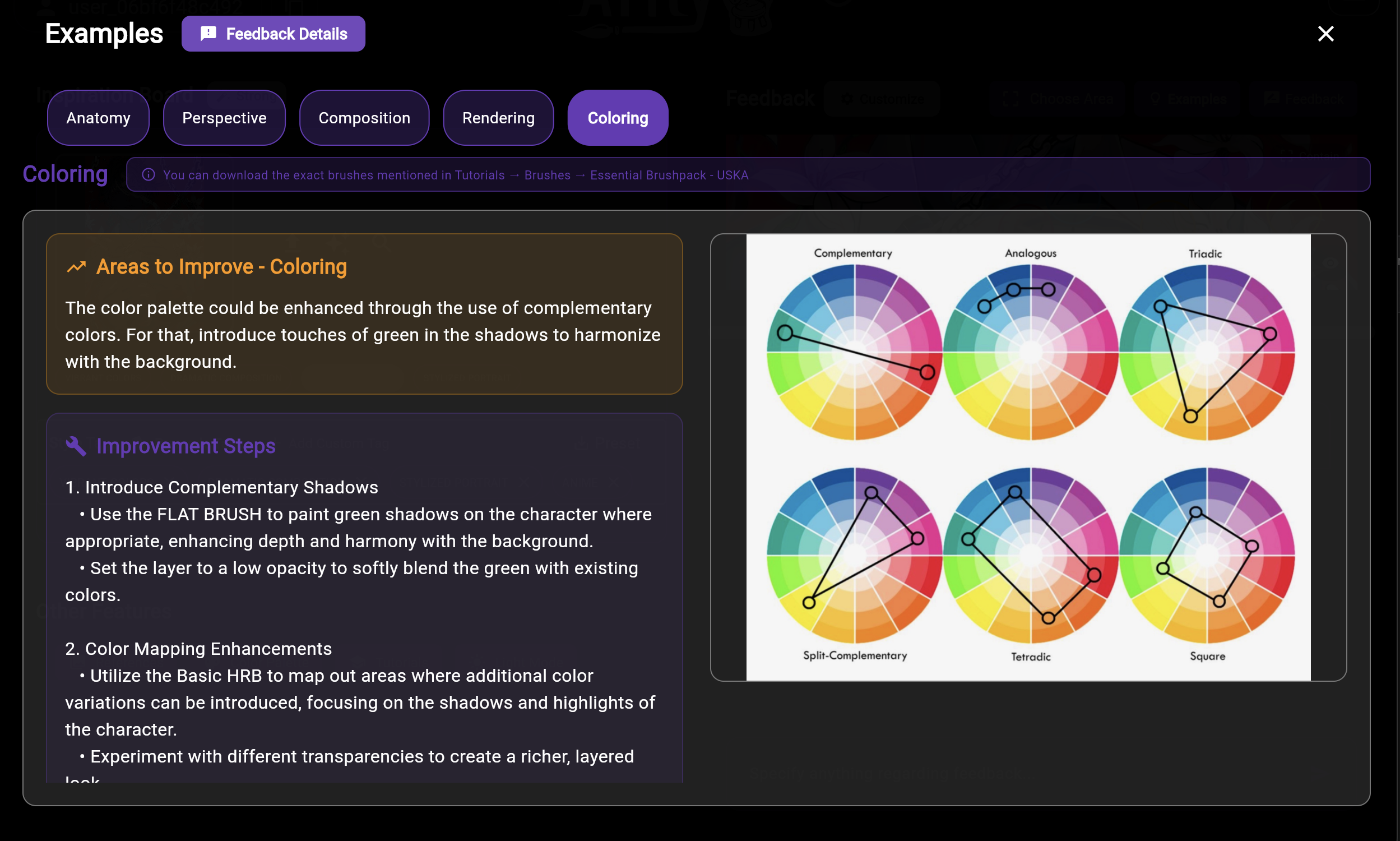}
    \caption{Coloring improvements.}
    \label{fig:feedback_coloring}
    \Description{A screenshot of the Examples view for the Coloring category. The left side provides text instructions under Areas to Improve - Coloring and Improvement Steps, advising the user to introduce complementary green shadows to harmonize with the background and providing steps for color mapping and brush usage. The right side shows a reference graphic of six color wheels, each illustrating a different color harmony theory: Complementary, Analogous, Triadic, Split-Complementary, Tetradic, and Square.}
  \end{subfigure}

  \caption{Example view for concrete improvements. This view expands the \textit{Areas to Improve} in the feedback details view. The improvement steps are generated with the examples for improvement prompt given in \autoref{sec:prompt_examples}.}
  \label{fig:combined_feedback}
\end{figure*}

Figures~\ref{fig:feedback_table}--\ref{fig:combined_feedback} show readable screenshots of examples for the three different feedback levels described in \autoref{sec:user_journey}.

\section{Participants Details}
\label{sec:participants_details}
\autoref{tab:participants} gives the full participant details as summarized in \autoref{sec:participants}.
\begin{table*}[h!]
    \centering
    \caption{Participant details. \textit{YOE} stands for years of experience. \textit{Gen.~Used} means whether the participant used the reference generation feature.}
    \label{tab:participants}
    \Description{A table detailing the demographics, prior AI experience, and study condition assignments for 38 participants. The table consists of eight columns: Participant ID, Age group, Years of Experience (YOE), self-reported Expertise, previously used AI Tools, level of Concern About AI in Art, Study Condition (Weak or Strong), and whether Reference Generation was Used (Yes or No). The data shows that the majority of participants are aged 18 to 24, identify as Intermediate or Advanced artists, and have 3 or more years of digital art experience. Many participants report prior use of ChatGPT, while a notable portion report no prior AI tool usage. Participant concerns regarding AI in art vary widely, ranging from 'Not at all concerned' to 'Very concerned'. The table lists 18 participants in the Weak condition (none of whom used reference generation) and 20 participants in the Strong condition (8 of whom used reference generation).}
    \begin{tabular}{lllclllc}
        \toprule
        ID & Age & YOE & Expertise & AI Tools & AI in Art Concern & Condition & Gen. Used \\
        \midrule
        W1 & 18-24 & $<$ 1 & Beginner & No & Neutral & Weak & No \\
        W2 & 18-24 & 5+ & Advanced & ChatGPT, PixAI & Not at all concerned & Weak & No \\
        W3 & 25-34 & 3-5 & Intermediate & No & Not concerned & Weak & No \\
        W4 & 18-24 & 5+ & Professional & ChatGPT & Not at all concerned & Weak & No \\
        W5 & 18-24 & 5+ & Intermediate & No & Not at all concerned & Weak & No \\
        W6 & $<$ 18 & 5+ & Intermediate & ChatGPT & Neutral & Weak & No \\
        W7 & 25-34 & 5+ & Intermediate & ChatGPT & Not concerned & Weak & No \\
        W8 & 18-24 & 3-5 & Intermediate & ChatGPT & Concerned & Weak & No \\
        W9 & 18-24 & 3-5 & Intermediate & No & Neutral & Weak & No \\
        W10 & 18-24 & 3-5 & Professional & No & Not concerned & Weak & No \\
        W11 & 25-34 & 5+ & Intermediate & No & Concerned & Weak & No \\
        W12 & 25-34 & 3-5 & Beginner & ChatGPT & Concerned & Weak & No \\
        W13 & 18-24 & 1-2 & Intermediate & ChatGPT, DeepL & Neutral & Weak & No \\
        W14 & 18-24 & 1-2 & Beginner & ChatGPT & Concerned & Weak & No \\
        W15 & 18-24 & 3-5 & Intermediate & Chat GPT & Not at all concerned & Weak & No \\
        W16 & 18-24 & 3-5 & Beginner & No & Neutral & Weak & No \\
        W17 & 18-24 & 3-5 & Intermediate & No & Neutral & Weak & No \\
        W18 & 18-24 & 1-2 & Beginner & No & Very concerned & Weak & No \\
        S-NR-1 & 18-24 & 5+ & Advanced & C.AI & Very concerned & Strong & No \\
        S-NR-2 & 18-24 & 3-5 & Intermediate & No & Concerned & Strong & No \\
        S-NR-3 & 18-24 & 3-5 & Intermediate & Dream.ai & Neutral & Strong & No \\
        S-NR-4 & 25-34 & 5+ & Advanced & ChatGPT, Gemini & Concerned & Strong & No \\
        S-NR-5 & 18-24 & 5+ & Advanced & ChatGPT, C.AI & Very concerned & Strong & No \\
        S-NR-6 & 18-24 & 5+ & Advanced & ChatGPT & Very concerned & Strong & No \\
        S-NR-7 & 18-24 & $<$ 1 & Intermediate & No & Not concerned & Strong & No \\
        S-NR-8 & 35+ & 5+ & Professional & ChatGPT & Concerned & Strong & No \\
        S-NR-9 & 18-24 & 5+ & Intermediate & No & Very concerned & Strong & No \\
        S-NR-10 & 35+ & 5+ & Advanced & ChatGPT & Very concerned & Strong & No \\
        S-NR-11 & 18-24 & 3-5 & Intermediate & ChatGPT, Copilot & Concerned & Strong & No \\
        S-NR-12 & 18-24 & 1-2 & Intermediate & No & Concerned & Strong & No \\
        S-R-1 & 18-24 & 3-5 & Advanced & ChatGPT, Copilot & Concerned & Strong & Yes \\
        S-R-2 & 18-24 & 1-2 & Intermediate & ChatGPT, Perplexity & Concerned & Strong & Yes \\
        S-R-3 & 18-24 & 3-5 & Intermediate & ChatGPT & Very concerned & Strong & Yes \\
        S-R-4 & 18-24 & 5+ & Intermediate & ChatGPT & Very concerned & Strong & Yes \\
        S-R-5 & 18-24 & 3-5 & Intermediate & ChatGPT & Not concerned & Strong & Yes \\
        S-R-6 & 18-24 & $<$ 1 & Beginner & No & Concerned & Strong & Yes \\
        S-R-7 & 18-24 & 5+ & Advanced & Pixai, Midjourney, ChatGPT & Concerned & Strong & Yes \\
        S-R-8 & 18-24 & 5+ & Advanced & ChatGPT & Concerned & Strong & Yes \\
        \bottomrule
    \end{tabular}
\end{table*}

\section{Full Questionnaire Items}
\label{sec:full_qestionnaire}
\autoref{tab:questionnaire} gives the full questionnaire items described in \autoref{sec:questionnaire_design}.
\begin{table*}[h!]
  \centering
  \caption{Post-task questionnaire items. All Likert-scale items were five-point, with 1~=~Strongly Disagree, 2~=~Disagree, 3~=~Neutral, 4~=~Agree, 5~=~Strongly Agree.}
  \label{tab:questionnaire}
  \Description{The table is organized into two sections: Likert-scale items and open-ended questions. The Likert-scale items used a five-point scale from 1 (Strongly Disagree) to 5 (Strongly Agree) and include nine metrics. 1. Agency: I felt the AI helped me without its suggestions overpowering my style. 2. Control: I felt full control over Artly's suggestions and could change the feedback to my liking. 3. Style Adaptation: Artly adapted well to my specific individual style, including recognition of selected style tags. 4. Feeling Creative: I feel more creative after using Artly. 5. New Ideas: Artly helped me to come up with new ideas and directions in my work. 6. Self-Improvement: I would say that Artly helped me to improve my illustration. 7. Discomfort: I felt discomfort about AI being part of my workflow. 8. Non-AI Features More Helpful: Non AI features like color palettes, videos, guides, and brushes were more helpful than AI Features like custom feedback, suggested tags, and reference generation. 9. Attitude Change: After using Artly, my attitude towards AI as a supporting tool in the Art industry has changed. The open-ended questions section includes three items. 1. Attitude Change: Why did your attitude change or not change?. 2. Intention to Use: Would you use Artly again for future projects? Why or why not?. 3. Learning: What is something new you learned or applied to your Illustration?.}
  \renewcommand{\arraystretch}{1.4} 
  
  \begin{tabular}{p{2.4cm} p{11.2cm}} 
    \toprule
    \textbf{Name} & \textbf{Question} \\
    \midrule
    \multicolumn{2}{c}{\textit{Likert-scale items}} \\
    \midrule
    Agency           & I felt the AI helped me without its suggestions overpowering my style. \\
    Control          & I felt full control over \textit{Artly}'s suggestions and could change the feedback to my liking. \\
    Style Adaptation & \textit{Artly} adapted well to my specific individual style (recognition of selected style tags). \\
    Feeling Creative & I feel more creative after using \textit{Artly}. \\
    New Ideas        & \textit{Artly} helped me to come up with new ideas and directions in my work. \\
    Self-Improvement & I would say that \textit{Artly} helped me to improve my illustration. \\
    Discomfort       & I felt discomfort about AI being part of my workflow. \\
    Non-AI Features More Helpful & Non AI features (color palettes, videos, guides, brushes) were more helpful for me than AI Features (custom feedback, suggested tags, reference generation). \\
    Attitude Change  & After using \textit{Artly}, my attitude towards AI as a supporting tool in the Art industry has changed. \\
    \midrule
    \multicolumn{2}{c}{\textit{Open-ended questions}} \\
    \midrule
    Attitude Change  & Why did your attitude change or not change? \\
    Intention to Use & Would you use \textit{Artly} again for future projects? Why or why not? \\
    Learning         & What is something new you learned/applied to your Illustration? \\
    \bottomrule
  \end{tabular}
\end{table*}

\end{document}